\documentclass[12pt]{article}

\usepackage[T1]{fontenc}
\usepackage{amsmath,amssymb,amsthm,amsfonts}
\usepackage{mathtools}
\usepackage{graphicx,epsfig,float,rotating}
\usepackage{caption}
\usepackage{subfigure}
\usepackage{float}
\usepackage{booktabs,multirow,array,threeparttable}
\usepackage{enumerate}
\usepackage{nicefrac,eurosym}
\usepackage{setspace}
\usepackage{xcolor}
\usepackage{soul}
\usepackage{sgame}
\usepackage{tikz}
\usetikzlibrary{arrows.meta,positioning,bending}
\usepackage[textwidth=30mm]{todonotes}
\usepackage{natbib}
\usepackage{geometry}
\usepackage{hyperref}
\usepackage{kpfonts}
\usepackage{enumitem}

\setlist[enumerate,1]{label=(\roman*)}
\setlist[enumerate,2]{label=(\alph*)}

\allowdisplaybreaks
\hypersetup{colorlinks=true,linkcolor=blue,citecolor=blue,urlcolor=blue}

\bibpunct[: ]{(}{)}{;}{a}{}{,}

\newtheorem{theorem}{Theorem}
\newtheorem{lemma}{Lemma}

\newtheorem{claim}{Claim}

\newtheorem{corollary}{Corollary}
\newtheorem{ass}{Assumption}

\newtheorem{definition}{Definition}
\newtheorem{example}{Example}
\newtheorem{observation}{Observation}

\newtheorem{proposition}{Proposition}

\title{Memoryless Algorithmic Collusion: \\ Sure to Fail, Slow to Fall\thanks{This paper was previously circulated as ``On Mechanism underlying Algorithmic Collusion''. We acknowledge Giacomo Calzolari, Jean-Edouard Colliard, Zhiwei Cui, Arthur Dolgopolov, Satoru Takahashi and Tristan Tomala for their helpful comments and discussions. We also acknowledge seminar and conference participants at GAIMSS 2024, Stony Brook Game Theory Conference 2024, GAMES 2024, Youth Forum of Digital Economics 2026, BEAT 2026, China Academy of Science, Fudan University and Shandong University for helpful comments. All errors are our own. }}
\author{Zhang Xu\thanks{School of Economics, Renmin University of China. Email: \href{mailto:xuzhang@ruc.edu.cn}{xuzhang@ruc.edu.cn}} \and Wei Zhao\thanks{
School of Economics and Management, Tsinghua University. Email: \href{mailto:wei.zhao@outlook.fr}{wei.zhao@outlook.fr}}}
\date{This Version: August 8, 2026\\[5pt] Latest Version: \href{https://365.kdocs.cn/l/ce8xU2qp9eOl}{Click Here}}

\begin{document}

\maketitle

\begin{abstract}
\noindent This paper shows that, in a class of Bertrand-style competition games, memoryless Q-learning algorithms should adapt to Nash Equilibrium given sufficient explorations in the long-run. This is also verified through accelerating simulations, while the convergence time grows super-exponentially for high discount factors. The resilience of collusive outcomes in the short-run is due to exploration of unprofitable actions, making the system resemble random walk. A structural model is proposed to estimate the convergence time, which not only explains the role of discount factor, but also unveils the non-monotonic relation in learning rate, which is overlooked in the literature. \\

\end{abstract}

\newpage
{
  \hypersetup{linkcolor=black}
  \tableofcontents
}

\clearpage
\pagenumbering{arabic}
\setcounter{page}{1}

\newpage

\section{Introduction}

The development of artificial intelligence (AI) is leading to an increasing number of decisions being delegated to AI. One prominent example is algorithmic pricing,\footnote{Recent evidence can be found in \cite{Chen2016}, \cite{brown2023competition} and \cite{CalderWang2024}.} which can respond rapidly to changing market conditions and process large amounts of consumer data.\footnote{See \cite{Bertini2021} and \cite{spann2024algorithmic}.} At the same time, concerns about algorithmic collusion have grown,\footnote{See \cite{oecd2017algorithmsandcollusion, oecd2018personalised}.} as such behavior has been documented in simulations and received empirical support.\footnote{In simulations, \cite{banchio2022artificial}, \cite{banchioartificial}, \cite{colliard2022algorithmic}, \cite{xu2024artificial}, and \cite{Abada2024} document memoryless algorithmic collusion, while \cite{Calvano2020}, \cite{klein2021autonomous}, \cite{dou2025ai}, and \cite{johnson2023platform} document algorithmic collusion with memory. In empirics, \cite{assad2024algorithmic} provides indirect evidence for algorithmic collusion.} Despite these findings, there is no consensus on the mechanisms behind algorithmic collusion in simulations. One view holds that supra-competitive pricing is a stable long-run outcome; however, this literature often relies on limiting cases, simplified environments, and restrictive assumptions.\footnote{For instance, \cite{banchioartificial} considers the limiting case where the learning rate converges to zero, which allows the discrete stochastic process to be approximated by continuous-time deterministic ordinary differential equations, and focuses only on symmetric scenarios. \cite{lambin2024less} approximates the learning process by dividing it into two phases, one with full exploration and the other without any exploration.} An alternative view suggests that algorithmic collusion stems from insufficient learning due to limited exploration.\footnote{\cite{asker2023impact} shows that the system can get stuck at any symmetric action profile without exploration, for various initial Q-values. \cite{colliard2022algorithmic} attributes collusive outcomes to the failure to learn the Nash equilibrium and shows that additional uncertainty further impedes learning.} This perspective, however, leaves two questions open. First, it is unclear whether sufficient exploration guarantees convergence to the Nash equilibrium. Second, the severe delay in convergence remains unexplained, since simulations in the literature already allow substantial exploration. Moreover, although prior research shows that simulation outcomes are highly sensitive to parameter specifications,\footnote{For example, numerous simulation studies show that collusive outcomes cannot be sustained when the discount factor is small \citep[e.g.,][]{lambin2024less}.} both perspectives remain largely silent on how these parameters shape the learning dynamics.

This paper addresses these gaps. Focusing on multi-agent memoryless Q-learning in Bertrand-style games, we present two main results. First, we show, both \emph{theoretically} and \emph{in simulations}, that only the Nash equilibrium is stable in the long run. Second, we develop and estimate a simple structural model using simulation data to characterize the convergence time. The fitted model explains why memoryless algorithmic collusion persists in the short run and how the parameters shape the convergence speed.  

The class of Bertrand-style games we study includes the Prisoner’s Dilemma, Bertrand competition, and any nonlinear mixture of first- and second-price auctions. These games are played by a general family of memoryless reinforcement learning algorithms that includes Q-learning. Each agent maintains a Q-value for each action, encoding its estimated discounted payoff. She typically chooses the action with the highest Q-value (the greedy policy) but occasionally explores other actions according to a predetermined rule. Upon receiving a payoff, she updates \emph{only} the Q-value of the chosen action toward the sum of the stage payoff and the discounted maximal Q-value, where the latter serves as the estimate of future payoffs under the greedy policy.

The theoretical analysis proceeds in two steps. First, we shut down exploration, so the pair of Q-vectors evolves as an \emph{unperturbed} Markov process. We show that all recurrent states are absorbing, and that any symmetric action profile can be played forever at some absorbing state. Thus, any collusive outcome may persist when exploration vanishes. This also explains why action cycles do not emerge in simulations. Second, we ask which absorbing states survive persistent but vanishing perturbations. To this end, we let agents explore with a constant probability, which induces a \emph{perturbed} Markov chain with a unique stationary distribution. As the exploration probability converges to zero, so does the stationary distribution to its limit. A state is \textbf{\textit{stochastically stable}} \citep{Foster1990} if it remains in the support of the limiting distribution. We explicitly construct an absorbing state, termed the \textbf{\textit{Nash state}}, in which each agent chooses the Nash action under greedy policy and which can withstand a single mutation. We show that \emph{only} the Nash state is stochastically stable. In other words, memoryless algorithmic collusion dissolves with probability one in the long run, given sufficient exploration.

The core of the proof is to show that every absorbing state is connected to the Nash state through a chain of absorbing states, where each transition is induced by a single mutation. We establish this by constructing a \emph{strict partial order} over absorbing states, with the Nash state as a minimum element. Under this order, every other absorbing state can reach a strictly lower state through a single mutation. Based on the common wisdom that alternating undercutting leads to the Nash equilibrium, a natural order is supposed to rank states by the level of their greedy actions. This intuition presumes that the winner's action is always reinforced, which fails under Q-learning. The Q-values of low actions are \emph{inflated}, whether by optimistic initialization or by exploration of unprofitable actions, so a winner may earn too little to justify its inflated Q-value. The undercutting is then interrupted, as both agents rebound to higher actions. Another round of undercutting follows, and the action path resembles an Edgeworth cycle. A purely action-based order therefore cannot work. Our order instead tracks the Q-values beneath the action cycle. The key observation is this: whenever a bilateral rebound occurs, some low action must have been chosen and later abandoned; consequently, its Q-value strictly decreases. Thus, action cycles do \emph{not} result in Q-value cycles. Each round of the cycle gradually erodes the inflated Q-values, and the dynamics eventually converge to the Nash equilibrium.


By accelerating the simulations, we verify our theory computationally. However, convergence time grows super-exponentially with the discount factor, raising a critical question of what sustains collusive outcomes in the short run. The answer is that exploration recreates Q-value inflation: after a bilateral rebound depresses the Q-values of lower actions, exploration of these actions inflates them again. The dynamics therefore resemble a \emph{random walk} with the Nash state as an absorbing boundary. If we shut down this channel by preventing at least one player from exploring unprofitable actions, the system converges to the Nash equilibrium within a tractable timeframe. Therefore, we conclude that random-walk dynamics, rather than action cycles, underlie the short-run persistence of memoryless algorithmic collusion.

To bridge the gap between the long-run theoretical result and the short-run simulation outcomes, we approximate the Markov chain by a biased random walk. The random walk is characterized by its length and transition probabilities, with parameters estimated from simulation data. The drift, defined as the net forward probability toward the Nash state, determines the expected convergence time. The time is \emph{linear} in the length if the drift is positive, but \emph{exponential} if it is negative. This model, while parsimonious, provides a close fit to the current dataset and strong out-of-sample predictive performance. Under random exploration by both agents, the drift is almost always negative, while prohibiting at least one agent from exploring unprofitable actions makes it positive. Convergence time is monotonic in the discount factor, as a higher discount factor increases the length but reduces the drift. The effect of the learning rate is more subtle. Since the learning rate shortens the length but also reduces the drift, convergence time first decreases, then increases, and eventually decreases again when the discount factor is high. This \emph{non-monotonic} effect has been overlooked in the previous literature.


Finally, we apply our framework to other environments, including the Prisoner’s Dilemma and auctions, and find similar patterns to those in Bertrand competition. Viewed through the lens of how the exploration mode shapes convergence time, the framework explains the contrast between first- and second-price auctions documented by \cite{banchio2022artificial}, as well as why sequential entry promotes competition \citep{lambin2024less}. For algorithm design, we show that synchronous updating makes the Nash equilibrium the unique recurrent class, implying that collusion cannot persist even without exploration \citep{asker2023impact}.

\paragraph{Related literature.}

The literature on algorithmic collusion can be broadly divided into two strands depending on whether algorithms retain memory of past actions. An early strand studies memoryless algorithms. \cite{waltman2008q} shows that memoryless Q-learning agents can sustain collusion in Cournot competition, and subsequent studies find similar outcomes in financial markets \citep{colliard2022algorithmic}, auctions \citep{banchio2022artificial}, market segmentation \citep{xu2024artificial}, and algorithm design \citep{asker2023impact}. However, the mechanism behind memoryless algorithmic collusion remains unclear. A later strand introduces memory of past actions, allowing collusion to be sustained through repeated-game incentives. The seminal paper by \cite{Calvano2020} shows that Q-learning agents using past prices as state variables can learn grim-trigger-like strategies and sustain collusion. Subsequent studies extend this finding to various environments, including sequential pricing \citep{klein2021autonomous}, financial markets \citep{dou2025ai}, platform governance \citep{johnson2023platform}, imperfect monitoring \citep{Calvano2021}, and human-algorithm interaction \citep{Werner2023}. \cite{Abada2025} and \cite{Calvano2026} offer a comprehensive survey on this literature. Our paper contributes to the first strand by developing a framework that explains how memoryless algorithmic collusion emerges and persists.

Beyond simulations, a growing literature studies algorithmic collusion theoretically. Since Q-learning dynamics are analytically intractable, existing approaches typically rely on approximations. For memoryless algorithms, \cite{banchioartificial} replaces stochastic discrete-time dynamics with deterministic continuous-time dynamics as the learning rate vanishes, and shows that cycles of cooperation and defection can arise in the Prisoner's Dilemma. We provide a complementary analysis by studying learning rates bounded away from zero and a more general class of games (see Section~\ref{sec:application} for a detailed comparison). Approximating learning dynamics with purely random exploration followed by no exploration, \cite{lambin2024less} shows convergence to supra-competitive outcomes. His results can be interpreted as convergence to some collusive absorbing state where Q-values are initialized by pure exploration. This is therefore consistent with our Proposition~\ref{prp:absorbing} that such states can exist without exploration. For algorithms with memory, \cite{askenazi2024bounds} establishes a folk-theorem-style result under a stochastic approximation framework. Their convergence is local, requiring specific initial conditions and sufficiently small learning rates; ours is global, holding from any initialization under persistent exploration. Other studies consider alternative learning protocols, such as self-play Q-learning \citep{bertrand2025self} and multi-armed bandit algorithms \citep{hansen2021frontiers}.

Instead of relying on approximations, our paper analyzes the underlying discrete stochastic process directly through finite Markov chains and identifies the long-run outcome through stochastic stability. \cite{waltman2007theoretical} first applied stochastic stability to algorithmic collusion in the Prisoner’s Dilemma, but restricted attention to a learning rate of one and undiscounted payoffs; \cite{dolgopolov2024reinforcement} extends the analysis to arbitrary learning rates. We further extend this approach to a class of Bertrand-style games with multiple actions and positive discount factors. Our companion paper \citep{Xu2026} further extends to general competition games, including Cournot and smooth Bertrand competition. Unlike approximation-based approaches, this direct approach establishes that under persistent non-discriminatory exploration, such as the $\varepsilon$-greedy policy, the learning dynamics converge to the Nash equilibrium in the long run. Moreover, we provide a deeper analysis of the mechanism underlying the persistence of collusive outcomes in the short run.

Methodologically, our paper contributes to the literature on stochastic stability. The foundational works \citep{Foster1990, kandori1993learning, young1993evolution} use stochastically stable states to select among multiple equilibria; subsequent work studies the speed of convergence and intermediate outcomes \citep{ellison2000basins, cui2010escape, levine2016dynamics}. We apply stochastic stability to select among absorbing states, which need not be equilibria, of the Markov processes induced by multi-agent Q-learning. The existing theory of convergence speed, however, cannot distinguish among the Q-learning parameters: different parameter values yield the same speed measure under the methods in this literature. We therefore complement it with a structural estimation of convergence times.

\section{Model}\label{sec:preliminaries}

Consider a two-player $K \times K$ symmetric game. Each player $i \in I = \{1,2\}$  has the action set $A = \{1, 2, \ldots, K\}$. Let $u(k, k')$ denote a player's  payoff from action $k \in A$ when the opponent plays $k' \in A$.

\begin{ass}[Bertrand-style Competition Games]\label{ass:game}
    The payoff function $u(\cdot,\cdot)$ satisfies:
    \begin{enumerate}
    \item \textit{(ranked actions)} $\underline{u} < u(1,1) < u(2,2) < \cdots < u(K,K)$;
    \label{assumption:1}
    \item \textit{(uniform losing payoff)} $u(k, k') = \underline{u}$ for all $k' < k$;
    \label{assumption:2}
    \item \textit{(profitable undercut)} for every $k > 1$, there exists $k' < k$ such that $u(k', k) > u(k, k)$;\label{assumption:3}
    \item \textit{(monotonicity in the opponent's action)} for every $k$, $u(k, k')$ is nondecreasing in $k'$.\label{assumption:4}
\end{enumerate}
\end{ass}

In this class of Bertrand-style games, $(1,1)$ is the unique Nash equilibrium (NE), and it is strict. The ordered action space captures the essence of  collusion. Assumption~\ref{ass:game}.\ref{assumption:1} ranks the actions:  action $1$ is the ``Nash action,'' yielding the lowest payoff among all symmetric  profiles, and higher symmetric profiles yield higher payoffs. Convergence to a  higher action thus represents a greater degree of collusion. Accordingly, we say that algorithmic collusion arises if a learning algorithm attains expected payoffs  strictly above the strict NE payoff.  Assumption~\ref{ass:game}.\ref{assumption:2} states that a player choosing a higher action than the opponent loses and receives only the outside option  $\underline{u}$. Assumption~\ref{ass:game}.\ref{assumption:3} states that at any symmetric profile above the Nash action, some lower action is a profitable deviation; hence collusive outcomes are not self-enforcing. Assumption~\ref{ass:game}.\ref{assumption:4} states that, fixing a player's own action, a higher action by the opponent never hurts the player.

\begin{example}[Prisoner's Dilemma]\label{eg:pd}
    Each player either defects ($D$) or cooperates ($C$), and the payoffs satisfy the standard ranking $u(C,D) < u(D,D) < u(C,C) < u(D,C)$. Let action $1$ denote $D$ and action $2$ denote $C$, and let  $\underline{u} = u(C,D)$. Then Assumption~\ref{ass:game} holds.
\end{example}

\begin{example}[Bertrand Competition]\label{eg:bertrand}
    Two firms with common marginal cost $c$ face market demand $D(p)$ and simultaneously choose prices from a sufficiently fine grid $\{p_1, \ldots, p_K\}$ with $c < p_1 < \cdots < p_K < p^m$, where $p^m$ is the monopoly price, below which the profit $(p-c)D(p)$ is strictly increasing. The lower-priced firm serves the entire market; ties split demand equally. Assumption~\ref{ass:game} holds with $\underline{u} = 0$ and both firms setting the lowest price is the unique strict NE. 
\end{example}

\begin{example}[Generalized First- and Second-price Auction]\label{eg:auction}
    Two bidders with common value $v$ compete in a sealed-bid auction, choosing bids from a sufficiently fine grid $\{b_1, \ldots, b_K\}$ with $v > b_1 > b_2 > \cdots > b_K \geq 0$. The higher bidder wins; ties split the object equally. A winner who bids $b$ against an opposing bid $b'$ pays $p(b, b')$, where
    \[
    b' \leq p(b, b') \leq b \qquad \text{and} \qquad 
    p(b, b') \text{ is nondecreasing in } b'.
    \]
    This formulation nests the first price auction ($p = b$) and the second price auction ($p = b'$). In the unique strict NE, both bidders submit the highest bid. Assumption~\ref{ass:game} holds with $\underline{u} = 0$.\footnote{Assumptions~\ref{ass:game}.\ref{assumption:1} and~\ref{ass:game}.\ref{assumption:2} are immediate: symmetric payoffs $\frac{1}{2}(v - b_k)$ increase as bids decrease, and a losing bidder earns zero. For Assumption~\ref{ass:game}.\ref{assumption:3}, since $p(b_{k-1}, b_k) \leq b_{k-1}$, raising the bid from $b_k$ to $b_{k-1}$ yields $u(k-1,k) \geq v - b_{k-1} \geq \frac{1}{2}(v - b_k) = u(k,k)$ on a sufficiently fine grid. For Assumption~\ref{ass:game}.\ref{assumption:4}, a lower opposing bid weakly lowers the winner's payment and never worsens the player's winning status.}
\end{example}

Two model-free reinforcement learning algorithms repeatedly play a fixed game  in this class. With no knowledge of the underlying environment, each algorithm  interacts with it repeatedly and updates its payoff estimates based on  realized payoffs. Let $Q_i^a$ denote agent $i$'s estimate of the discounted  payoff from choosing action $a$, and let $\mathbf{Q}_i = (Q_i^a)_{a \in A}$ denote the associated vector. In each period, each player selects an action based only on her current Q-values and then updates them based only on the realized payoff and the current Q-values. The induced stochastic process is therefore a Markov chain with \textbf{\textit{state}} $s = (\mathbf{Q}_1, \mathbf{Q}_2)$; we write  $Q_i^a(s)$ for the Q-value of action $a$ at state $s$.

\paragraph{Updating rule.}
Suppose that at state $s_t$ player $i$ plays $a_i$ and her opponent plays  $a_{-i}$. Player $i$'s \textbf{\textit{estimated discounted payoff}} for action $a_i$ is
\begin{equation}\label{eq:target}
    V_i(a_i, a_{-i};\, s_t) 
    := u(a_i, a_{-i}) + \delta \max_{a \in A} Q_i^{a}(s_t),
\end{equation}
where $\delta \in [0,1)$ is the discount factor: the first term is the  realized payoff, and the second is the estimated continuation value under the perceived optimal action. After the profile $(a_1, a_2)$ is played, the state transitions to  $s_{t+1} = \Phi(s_t;\, a_1, a_2)$, where $\Phi$ moves the played action's  Q-value strictly toward its target, without overshooting, and leaves all  other Q-values unchanged. Formally, whenever $Q_i^{a_i}(s_t) \neq V_i(a_i, a_{-i}; s_t)$,
\begin{align}
    &\bigl| Q_i^{a_i}(s_{t+1}) - V_i(a_i, a_{-i}; s_t) \bigr| 
    < \bigl| Q_i^{a_i}(s_t) - V_i(a_i, a_{-i}; s_t) \bigr|, 
    \tag{closer}\label{upd:closer}\\
    &\bigl( Q_i^{a_i}(s_{t+1}) - V_i(a_i, a_{-i}; s_t) \bigr)
    \bigl( Q_i^{a_i}(s_t) - V_i(a_i, a_{-i}; s_t) \bigr) \geq 0;
    \tag{no-overshoot}\label{upd:no-overshoot}
\end{align}
otherwise $Q_i^{a_i}(s_{t+1}) = Q_i^{a_i}(s_t)$; and in either case,
\begin{equation}
    Q_i^{a}(s_{t+1}) = Q_i^{a}(s_t) 
    \qquad \text{for all } a \neq a_i.
    \tag{asynchronous}\label{upd:async}
\end{equation}
Condition \eqref{upd:async} contrasts with synchronous updating, as in best-response dynamics and no-regret learning: only the played action's Q-value is revised. Conditions 
\eqref{upd:closer}--\eqref{upd:async} generalize standard Q-learning,
\begin{equation}\label{eq:standardQ}
    Q_i^{a_i}(s_{t+1}) 
    = (1-\alpha_{i})\, Q_i^{a_i}(s_t) + \alpha_{i}\, V_i(a_i, a_{-i}; s_t),
\end{equation}
where $\alpha_{i} \in (0,1]$ is player $i$'s learning rate.

Treating Q-values as real numbers would require establishing convergence over  an infinite state space without yielding additional insight. Moreover, a  computer running a learning algorithm carries a built-in limit on the  floating-point representation of numbers \citep{goldberg1991every}. To account  for \textbf{\textit{finite machine precision}}, we assume that all Q-values lie on a  discrete grid $\mathcal{D} \subseteq \mathbb{R}$ that contains finitely many points in any bounded interval.\footnote{This discreteness allows us to invoke the machinery of finite Markov chains, in particular the notion of stochastic stability introduced by \cite{Foster1990}. However, it also affects the updating rule. When an updated Q-value falls off the grid, it should be rounded to a nearby grid point. Thus, convergence results hold only up to the grid resolution. In particular, any grid point within a neighborhood of fixed point of the continuous updating rule may serve as a fixed point of its discrete counterpart. Making this explicit would only complicate the exposition without adding insight. Consequently, the main text abstracts away these discrete mechanics. Update processes are treated as continuous. Nonetheless, Q-value convergence to the fixed point occurs in finite steps, not in the limit. Appendix~\ref{appendix:discrete_updating} gives the rigorous discrete formulation and shows that all results are unaffected for a sufficiently fine grid.} Throughout the paper, we restrict attention to the \textbf{\textit{finite state space}}
\begin{equation}
    \mathcal{S} = \left\{(\mathbf{Q}_1, \mathbf{Q}_2) :
    Q_i^a \in \left[\underline{Q}^a, \bar{Q}\right] \cap \mathcal{D},
    \ a\in A,\ i \in I \right\},
\end{equation}
where the lower bound is equal to the lowest discounted payoff attainable by that action, i.e., $\underline{Q}^a=u(1,1)/(1-\delta)$ for $a=1$ and $\underline{Q}^a=\underline{u}/(1-\delta)$ for $a\neq1$, and $\bar{Q} \geq \max_{a,a'} u(a,a')/(1-\delta)$ is an arbitrary upper bound. The lower bounds are without loss. Updates target discounted payoffs, so any $Q^a< \underline{Q}^a$ enters the interval in finitely many updates and never leaves. A state outside $\mathcal{S}$ only if actions with out-of-range Q-values are never played. Once these actions are explored, the state enters $\mathcal{S}$ permanently. We impose no further restriction on $\bar{Q}$, allowing for optimistic initialization, as in \cite{asker2023impact} and \cite{banchioartificial}.

\section{Learning Dynamics and Long-Run Behavior} \label{sec:stability}

In this section, we first study Q-learning without exploration and characterize recurrent classes of the resulting dynamics. We then introduce $\varepsilon$-greedy exploration, the simplest and most widely adopted exploration rule, under which every action is explored with equal probability. The main result is that, as exploration vanishes, the long-run probability of playing any profile other than the strict NE converges to zero. Supra-competitive outcomes in simulations are therefore short-run phenomena.

\subsection{Unperturbed Dynamics}

For each state $s$, an action maximizing $Q_i^a(s)$ is called a  \textbf{\textit{greedy action}} of player $i$. Let
\begin{equation}
    G_i(s) := \arg\max_{a \in A} Q_i^{a}(s), 
    \qquad 
    G(s) := G_1(s) \times G_2(s)
\end{equation}
denote player $i$'s greedy action set and the set of  greedy action profiles, and for any $S \subseteq \mathcal{S}$,  let $G_i(S) := \bigcup_{s \in S} G_i(s)$ and $G(S) := \bigcup_{s \in S} G(s)$. The decision rule is a  \textbf{\textit{greedy policy}}: it assigns positive probability to every greedy action and zero probability to all other actions. Joint with the updating rule, this greedy policy generate the set of \textbf{\textit{unperturbed processes}}
\begin{equation}\label{eq:unperturbed}
    \mathbb{P}_0 = \bigl\{ P_0 : P_0(s, s') > 0 
    \ \text{if and only if}\ 
    s' = \Phi(s;\, \mathbf{a}) \ \text{for some}\ \mathbf{a} \in G(s) \bigr\}.
\end{equation}

We are interested in the recurrent classes of the unperturbed dynamics. A  \textbf{\textit{recurrent class}} is a set of states that are mutually accessible and from which no outside state is accessible;\footnote{Formally,  given a transition matrix $P$, let $P^t$ denote its $t$-th power, so that $P^t(s, s')$ is the $t$-step transition probability from $s$ to $s'$. State $s'$ is  \emph{accessible} from $s$, written $s \to s'$, if $P^t(s, s') > 0$ for some  $t \geq 0$. A nonempty set $S \subseteq \mathcal{S}$ is a recurrent class if  (i) $s \to s'$ for all $s, s' \in S$, and (ii) no $s \in S$ and $s'' \notin S$ satisfy $s \to s''$.} $\mathcal{R}$ denotes the set of recurrent classes. 
When a recurrent class is a singleton $\{s\}$, the state $s$ is called an \textbf{\textit{absorbing state}}, and we abbreviate 
$\{s\} \in \mathcal{R}$ as $s \in \mathcal{R}$. The following proposition characterizes the recurrent classes of the Q-learning dynamics under the greedy policy.

\begin{proposition}\label{prp:absorbing}
    For any game satisfying Assumption~\ref{ass:game}, all recurrent classes are singletons. Moreover, $s \in \mathcal{R}$ if and only if there exists $a \in A$ such that, for each $i \in I$,
    \begin{equation}\label{eq:absorbing}
        Q_i^a(s) = \frac{u(a,a)}{1 - \delta}
        \qquad \text{and} \qquad
        Q_i^a(s) > Q_i^{a'}(s) \quad \text{for all } a' \neq a.
    \end{equation}
\end{proposition}

The proof of Proposition \ref{prp:absorbing} is mainly divided into two steps. In the first step, we argue that no asymmetric profile can be played in a recurrent class $S$. Otherwise, the loser's Q-value would be driven down toward $\underline{u}/(1-\delta)$, eventually making the Nash action greedy. Since then, both players' Q-values of action $1$ rise above $\underline{u}/(1-\delta)$ and stay there forever, contradicting the recurrence of original state. In the second step, we show that $S$ cannot contain two symmetric profiles. Otherwise, switching between them requires passing through an asymmetric profile, which is impossible by the first step. Accordingly, $G(S)$ is a single symmetric profile, and playing it repeatedly locks in the Q-values, making $S$ a singleton.

A key implication of Proposition~\ref{prp:absorbing} is that, from any initial Q-values, the unperturbed dynamics eventually settle on some symmetric action profile. Since this profile can be any symmetric one, the pair of Q-learning algorithms may get stuck at a supra-competitive outcome when exploration vanishes. This echoes \cite{asker2023impact}, who obtain supra-competitive outcomes in the absence of exploration, and \cite{Abada2024}, who show that supra-competitive profits can arise from imperfect exploration. Proposition~\ref{prp:absorbing} also explains why the memoryless algorithmic collusion observed in simulations exhibits such large variance. It further rules out cycles of action profiles in the long run.\footnote{\cite{dolgopolov2024reinforcement} states that if the game is solvable by iterated elimination of strictly dominated strategies (IESDS), then all recurrent classes are singletons. However, this is not true. The game in the following table is IESDS-solvable while there exists a recurrent set $S:= \{s: Q_1^{a_1}(s) \in [2,3], Q_1^{a_2}(s) = 1, Q_2^{b_1}(s) = Q_2^{b_2}(s) = 2 \text{ and } Q_2^{b_3}(s) = 1\}$, with $G_1(S) = \{a_1\}, G_2(S) = \{b_1, b_2\}$.

\begin{minipage}{0.9\linewidth}
\centering
\begin{tabular}{|c|c|c|c|}
\hline
& \(b_1\) & \(b_2\) & \(b_3\) \\
\hline
\(a_1\) & 3,2 & 2,2 & 2,3 \\
\hline
\(a_2\) & 1,0 & 1,0 & 1,1 \\
\hline
\end{tabular}
\end{minipage}}

\subsection{Perturbed Dynamics}\label{subsection:perturbed}

Reinforcement learning algorithms need to explore to learn their environment. Recent simulation studies typically adopt the \textbf{\textit{$\varepsilon$-greedy policy}}: with probability $1-\varepsilon$ the player follows the greedy policy, and with probability $\varepsilon$ she draws an action uniformly from $A$. Since the two players randomize independently, the probability of a one-step transition is of exactly the order $\varepsilon^k$, written $\Theta(\varepsilon^k)$, where $k \in \{0, 1, 2\}$ is the number of players taking non-greedy actions,\footnote{Here $f(\varepsilon) \in \Theta(\varepsilon^k)$ means that there exist constants $0 < m \leq M$ such that $m\,\varepsilon^k \leq f(\varepsilon) \leq M\,\varepsilon^k$ for all sufficiently small $\varepsilon$. To see the claim, note that player $i$ selects a greedy action $a \in G_i(s)$ with probability $(1-\varepsilon)\, g_i(a \mid s) + \varepsilon/|A|$, where $g_i(\cdot \mid s)$ is the distribution of the greedy policy, and selects each non-greedy action with probability $\varepsilon/|A|$. Consider a profile $(a_1, a_2)$. If both actions are greedy, it is played with probability
\[
    \left[(1-\varepsilon)\, g_1(a_1 \mid s) + \frac{\varepsilon}{|A|}\right]
    \left[(1-\varepsilon)\, g_2(a_2 \mid s) + \frac{\varepsilon}{|A|}\right]
    = \Theta(1);
\]
if exactly one action, say $a_2$, is non-greedy, with probability
\[
    \left[(1-\varepsilon)\, g_1(a_1 \mid s) + \frac{\varepsilon}{|A|}\right]
    \frac{\varepsilon}{|A|}
    = \Theta(\varepsilon);
\]
and if both are non-greedy, with probability $\varepsilon^2/|A|^2 = \Theta(\varepsilon^2)$. The transition probability $P_{\varepsilon}(s,s')$ sums these probabilities over all profiles with $\Phi(s;\, \mathbf{a}) = s'$ and is therefore of the order of the cheapest realizing profile. If no profile realizes $s'$, then $P_{\varepsilon}(s,s') = 0$ for all $\varepsilon$, corresponding to an infinite cost; by convention, $\Theta(\varepsilon^{\infty})$ denotes a probability of zero. Finally, $P_{\varepsilon}(s,s') \to P_0(s,s')$ as $\varepsilon \to 0$. The induced dynamics therefore satisfy Assumption~\ref{ass:perturbed_dynamics} below.} called \textbf{\textit{mutations}},\footnote{Strictly speaking, an \textbf{\textit{exploration}} draws a uniformly random action, which may happen to be greedy and hence need not produce a \textbf{\textit{mutation}}. Since the distinction is immaterial for our analysis, we use the two terms interchangeably.} in that step. This motivates the following assumption.

\begin{ass}\label{ass:perturbed_dynamics}
    For any $\varepsilon \in (0,1)$, the set of perturbed dynamics is
    \begin{equation}
        \mathbb{P}_{\varepsilon} := \bigl\{ P_{\varepsilon} : \lim_{\varepsilon \to 0} P_{\varepsilon}(s, s') = P_0(s,s'), \ P_{\varepsilon}(s, s') \in \Theta\bigl(\varepsilon^{c(s, s')}\bigr) 
    \bigr\},
    \end{equation}
    where the \textbf{\textit{one-step cost}} $c(s, s')$ is the minimal number of mutations required to move from $s$ to $s'$ in one step:
    \begin{enumerate}
        \item $c(s,s') = 0$ if $s' = \Phi(s;\, \mathbf{a})$ for some 
        $\mathbf{a} \in G(s)$;
        \item $c(s,s') = 1$ else if $s' = \Phi(s;\, a_i, a_{-i})$ for some 
        $i \in I$ and $(a_i, a_{-i}) \in G_i(s) \times 
        \bigl(A \setminus G_{-i}(s)\bigr)$;
        \item $c(s,s') = 2$ else if $s' = \Phi(s;\, a_0, a_1)$ for some 
        $(a_1, a_2) \in \bigl(A \setminus G_1(s)\bigr) \times 
        \bigl(A \setminus G_2(s)\bigr)$;
        \item $c(s,s') = \infty$ if $s' \neq \Phi(s;\, \mathbf{a})$ for all 
        $\mathbf{a} \in A \times A$.
    \end{enumerate}
\end{ass}

Note that this assumption is more general than the $\varepsilon$-greedy policy as it requires exploration to be neither uniform nor independent of Q-values.\footnote{Note that state-dependent factors do not alter the order $\varepsilon^{c(s,s')}$. Therefore, dynamics in $\mathbb{P}_{\varepsilon}$ can be induced by any exploration rule whose mutation probabilities are bounded above and below by constant multiples of $\varepsilon$. One typical example is the exploration rule where the probability of exploring action $a$ is $\beta(\mathbf{Q}_i) \varepsilon$ for some positive function $\beta$. However, the assumption rules out exploration rule that changes the order itself, like Boltzmann rule. Specifically, under Boltzmann exploration rule, the mutation probability for action $a$ is proportional to $e^{(Q_i^a - \max_{a'} Q_i^{a'})/\tau}$. Setting $\varepsilon := e^{-1/\tau}$, the one-step cost of this mutation becomes the Q-value gap $\max_{a'} Q_i^{a'} - Q_i^a$ rather than $1$. Escaping a collusive state, where competitive actions have deeply depressed Q-values, is then particularly costly. Consequently, this ``discriminatory'' exploration inherently favors collusive outcomes. Indeed, \cite{dolgopolov2024reinforcement} shows that under Boltzmann exploration, mutual cooperation in the Prisoner's Dilemma can be stochastically stable under certain conditions. Analyzing this exploration rule with more than two actions, however, is substantially more complex, and we exclude this channel from our analysis.} Appendix~\ref{app:proofs} shows that $P_\varepsilon$ has a unique stationary distribution, denoted by $\mu_\varepsilon$, and that $\mu_\varepsilon$ converges to a stationary distribution of $P_0$ as $\varepsilon\to0$. Following \cite{Foster1990}, a state $s$ is \textbf{\textit{stochastically stable}} if $\lim_{\varepsilon\to0}\mu_\varepsilon(s)>0$. This concept refines the notion of recurrent classes when the unperturbed dynamics has more than one recurrent class. As the exploration rate vanishes, the long-run probability of visiting any state that is not stochastically stable goes to zero.

Let $s^{NE}$ denote the \textbf{\textit{Nash state}}, defined by
\begin{equation}\label{eq:s_NE}
    Q_i^a(s^{NE}) = u(a, 1) + \frac{\delta\, u(1,1)}{1-\delta}, 
    \qquad \forall a \in A, \ \forall i \in I.
\end{equation}
By construction, for any $a > 1$,
\begin{equation*}
    Q_{i}^{a}(s^{NE}) = \underline{u} + \frac{\delta\, u(1,1)}{1-\delta}
    < \frac{u(1,1)}{1-\delta} = Q_{i}^{1}(s^{NE}) \qquad \text{for all } i\in I,
\end{equation*}
so at $s^{NE}$, both agents play the Nash action $1$ with probability one, i.e., $G(s^{NE}) = \{(1,1)\}$.


\begin{theorem}\label{thm:ss}
    The Nash state $s^{NE}$ is the unique stochastically stable state.
\end{theorem}

Theorem~\ref{thm:ss} shows that, with sufficient but arbitrarily small exploration, Q-learning algorithms adapt to the strict NE $(1,1)$ in the long run. The collusive outcomes observed in simulations are therefore a result of insufficient learning: as the exploration rate converges to zero, the long-run probability of collusive outcomes vanishes.

\paragraph{Proof Sketch.} Stochastic stability selects among absorbing states when the unperturbed dynamics admit more than one. Intuitively, the dynamics concentrate in the long run on states that are hard to leave and relatively easy to enter. By construction, it takes at least two mutations to leave the Nash state. It can be shown that, for any other absorbing states, there always exist a single mutation to escape. The core of the proof is to connect each absorbing state to the Nash state $s^{NE}$ through a chain of absorbing states, where each transition is induced by a single mutation. This is illustrated by Figure~\ref{fig:proof_intuition}. To establish this argument, we explicitly construct a strict partial order in the next section. This order exhibits two properties. First, $s^{NE}$ is a minimum element. Second, for any other absorbing state, there always exist a single mutation reaching another absorbing state ranked below.

\begin{figure}[!t]
\centering
\begin{tikzpicture}[
    >=Stealth,
    scale=1.2,
    absorbing/.style={
        circle,
        fill=black,
        inner sep=7pt
    },
    transient/.style={
        circle,
        draw=black,
        dashed,
        very thick,
        fill=white,
        inner sep=6pt
    },
    costlabel/.style={
        font=\small,
        inner sep=2pt,
        midway
    },
    every node/.style={font=\small}
]

\node[circle, draw, thick, minimum size=0.6cm] 
(sne) at (7,0) {$s^{NE}$};

\node[absorbing] (a1) at (0.0,0.8) {};
\node[absorbing] (a2) at (1.2,-1.2) {};
\node[absorbing] (a3) at (2.0,1.2) {};
\node[absorbing] (a4) at (2.7,-0.5) {};
\node[absorbing] (a5) at (3.7,0.7) {};
\node[absorbing] (a6) at (3.9,-1.4) {};
\node[absorbing] (a7) at (5.2,0.2) {};
\node[absorbing] (a8) at (5.5,-1.2) {};
\node[absorbing] (a9) at (6.0,1.6) {};

\draw[->, thick] (a1) -- node[costlabel, above] {$1$} (a3);
\draw[->, thick] (a2) -- node[costlabel, below] {$1$} (a4);
\draw[->, thick] (a3) -- node[costlabel, above] {$1$} (a5);
\draw[->, thick] (a4) -- node[costlabel, below] {$1$} (a5);
\draw[->, thick] (a5) -- node[costlabel, above] {$1$} (a7);
\draw[->, thick] (a6) -- node[costlabel, below] {$1$} (a8);
\draw[->, thick] (a7) -- node[costlabel, above] {$1$} (sne);
\draw[->, thick] (a8) -- node[costlabel, below] {$1$} (sne);
\draw[->, thick] (a9) -- node[costlabel, right] {$1$} (sne);

\node[transient] (t1) at (-1.0,2.1) {};
\node[transient] (t2) at (0.7,2.5) {};
\node[transient] (t3) at (3.0,2.5) {};
\node[transient] (t4) at (-0.5,-1.3) {};

\draw[->, dashed, thick] 
(t1) -- node[costlabel, left] {$0$} (a1);

\draw[->, dashed, thick] 
(t2) -- node[costlabel, right] {$0$} (a1);

\draw[->, dashed, thick] 
(t3) -- node[costlabel, right] {$0$} (a3);

\draw[->, dashed, thick] 
(t4) -- node[costlabel, above] {$0$} (a2);

\node (q) at (9.0,0) {\Large $?$};

\draw[->, thick] 
(sne) -- node[costlabel, above] {$\geq 2$} (q);

\begin{scope}[shift={(0.5,-2.5)}]

\node[absorbing] (la) at (0,0) {};
\node[right=8pt of la] {absorbing state};

\node[transient] (lt) at (4.2,0) {};
\node[right=8pt of lt] {transient state};

\end{scope}

\end{tikzpicture}

\caption{
Every absorbing state reaches $s^{NE}$ through a chain of
single-mutation transitions (cost $1$), while transient states flow in
without mutations (cost $0$). Escaping $s^{NE}$ requires at least two
mutations (cost $\geq 2$).
}
\label{fig:proof_intuition}

\end{figure}

\section{The Nash-Proximity Order}\label{sec:sketch}

The previous section outlined the proof of Theorem~\ref{thm:ss}. This section focuses on the key object in that proof: the order used to compare absorbing states. The first subsection gives the formal construction, and the second subsection explains the intuition behind the order and why it captures the way the learning dynamics move toward the Nash state. The construction of the order brings out a key economic insight: exploration plays a double-edged role in the convergence to the Nash state.

\subsection{Construction of the Order}

Recall from Proposition~\ref{prp:absorbing} that every absorbing state $s$ has a unique symmetric greedy profile $(a,a)\in G(s)$. Let $a(s)$ denote the action played by both players at $s$.

\begin{definition}[Nash-Proximity Order]\label{def:order}
Fix a mapping $\psi:A\to A$. Define a relation $\prec_\psi$ on $\mathcal R$ as follows: $s\prec_\psi s'$ if one of the following conditions holds:
\begin{enumerate}
\item\label{order_condition_1}
there exists $\hat a<\min\{a(s),a(s')\}$ such that $Q_1^a(s)=Q_1^a(s')$ for all $a<\hat a$ and $Q_1^{\hat a}(s)<Q_1^{\hat a}(s')$;
\item\label{order_condition_2}
condition~\ref{order_condition_1} fails and $a(s)<a(s')$;
\item\label{order_condition_3}
condition~\ref{order_condition_1} fails, $a(s)=a(s') > 1$, and  $Q_2^{\psi(a(s))}(s)>Q_2^{\psi(a(s'))}(s')$.
\end{enumerate}
\end{definition}

The order ranks absorbing states by their proximity to the NE through a lexicographic comparison of three features. It first compares the Q-values of the lowest actions; if these do not separate the two states, it compares their symmetric greedy actions; and if this too fails, it compares the Q-values of the mutation targets $\psi(a(s))$. The minimal elements of $\prec_\psi$ are exactly the absorbing states at which both players play the Nash action, i.e., those with $a(s) = 1$: any such state $s$ satisfies $s \prec_\psi s'$ for every absorbing state $s'$ with $a(s') \neq 1$, and no absorbing state is strictly below it.

Before illustrating this order, the following lemma lists two key properties. 
\begin{lemma}\label{lem:property-order}
    For any mapping $\psi:A\to A$, the relation $\prec_\psi$ is a strict partial order.\footnote{A strict partial order is irreflexive ($s\not\prec_\psi s$) and transitive ($s\prec_\psi s'$ and $s'\prec_\psi s''$ imply $s\prec_\psi s''$).} There exists a mapping $\psi: A \to A$ such that for every $s \in \mathcal{R}$ with $a(s) > 1$, there is $s' \in \mathcal{R}$ with $s' \prec_\psi s$ and $C(s, s') = 1$.
\end{lemma}

The first property shows that the order $\prec_\psi$ is well-defined. The second property ensures that a single mutation always moves the dynamics strictly closer to the NE, and the mapping $\psi$ specifies which lower action serves as the mutation target. Hence, every absorbing state is connected to a minimal element by a chain of single mutations. Once the dynamics reach a Nash-action absorbing state, the Q-value of the Nash action converges to the discounted Nash payoff; then,  condition~\eqref{upd:closer} drives the remaining Q-values to their values at $s^{NE}$.

It is worth noting that the comparison in condition \ref{order_condition_1} and \ref{order_condition_3} is asymmetric across players. When comparing the Q-values associated with lower actions, only player $1$'s Q-values are considered, whereas the comparison of the mutation target action is based on player $2$'s Q-values. This asymmetry stems from the distinct roles the two players play, which we illustrate below.

\subsection{Interpretation of the Order}

To interpretate the strict partial order, we draw a common wisdom similar to best response learning dynamics, based on which the convergence to the NE seems obvious. 

Suppose both Q-learning agents are stuck at some symmetric supra-competitive price at beginning, one agent (agent 1) explores through undercutting the price. Such practice makes him win the market game and the generated stage payoff then increases the Q-value of the undercut price. As a result, the undercut price will be reinforced and gradually, agent 1 will transit to the undercut price. Condition \ref{order_condition_3} exactly captures this process. Given that agent 1 firmly chooses the undercut price, agent 2 constantly loses the market game and his Q-value of the original price strictly decreases. When the Q-value is lower than the second highest one, agent 2 is supposed to change his action firmly. Such process will continue until agent 2 chooses some prices strictly lower than agent 1's. Now it is agent 1 whose Q-value strictly decreases, forcing him to change for another price. Such process, termed as \textbf{\textit{alternating undercut}}, is exactly captured by condition \ref{order_condition_2}. If this process can be iterated without interruption, then the dynamic system should gradually converge to the NE.




The key mechanism underlying the common wisdom is that the winner's chosen prices will always be reinforced. However, such mechanism does not always hold in the dynamics of Q-leaning. In the simulation, it is observed that Q-values of low prices are inflated. Therefore, if winning Q-agent undercut by charging sufficient low price, then the winner's Q-value (of this low price) is supposed to decrease instead. Given that, not only the loser, but also the winner have to rebound to some higher prices, termed as \textit{\textbf{bilateral rebound}}. Hence, the alternating undercut process will be interrupted, followed by a new round of alternating undercut process, making the price curves like Edgeworth cycle. When they coincidentally rebound to the same price high enough and the exploration rate vanishes at the same time, then they can get stuck at this supra-competitive price, as the Q-values of other prices are low and almost never updated.\footnote{Similar cyclical behavior is documented by \cite{banchioartificial}, who interpret it as the discrete-time counterpart of the pseudo-steady state on the boundary between the cooperation and defection regions in their continuous-time approximation.} 
\begin{figure}[t]
    \centering
    \includegraphics[width=\textwidth]{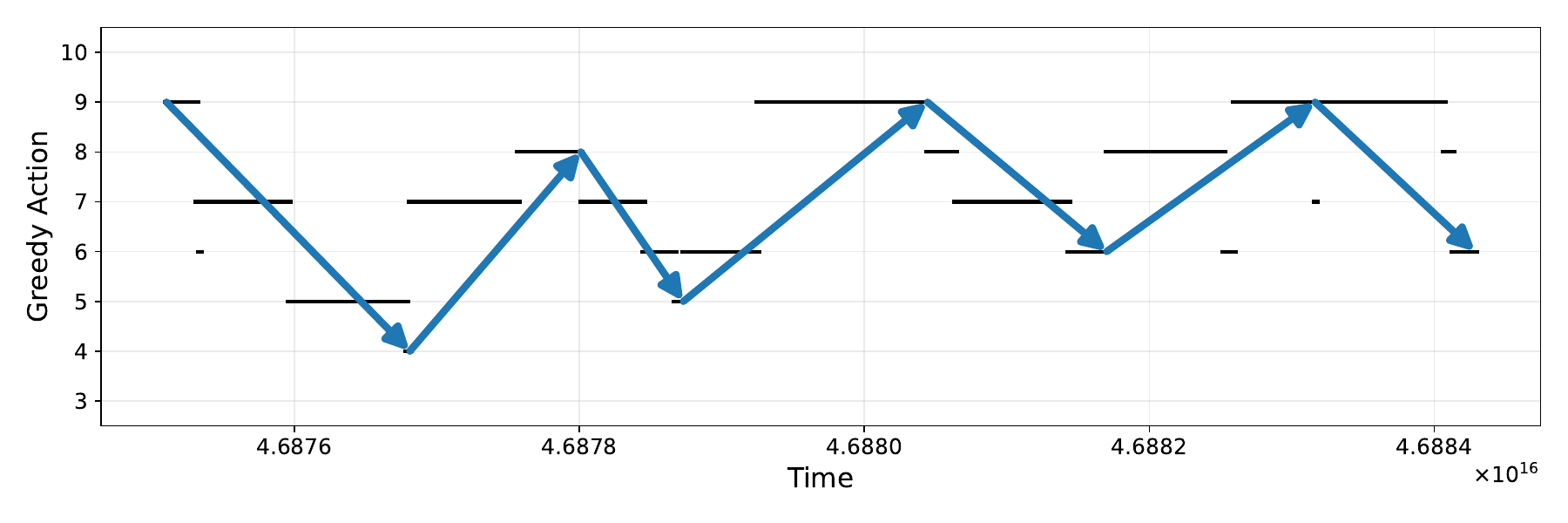}
    \caption{A representative sample of alternating undercuts and bilateral rebounds. The sample is generated under a symmetric Bertrand competition with zero marginal cost and $10$ actions evenly spaced between $0.2$ and $1$, with $\alpha = 0.15$, $\delta = 0.9$, and $\varepsilon = 10^{-10}$.}

    \label{fig:order_dynamics}
\end{figure}

The main source of Q-value inflation comes from exploration.\footnote{The other source of inflation comes from optimistic initialization \citep[cf.][]{asker2023impact, banchioartificial}, setting initial Q-values above the highest attainable discounted payoff.} To estimate the future payoff under greedy policy, the maximal Q-value is injected into each updated Q-values. If the action is chosen by exploration and unprofitable, i.e. $u(a_i,a_{-i}) < \max_{a'}Q_i^{a'}$, then the updating rule does inflate $Q_i^{a_i}$. Under the classical setting with time-decaying exploration studied in the literature, exploration is extensive in the early stages. Consequently, such a bubble always exists even if Q-values are initialized at a reasonable level.

Given cyclical price dynamics, the convergence to NE remains questionable. The key point is that \textbf{\textit{cycles in actions need not imply cycles in Q-values}}. A price-based order captures only the observed price path and misses the more important dynamics in the Q-values.

\begin{observation}\label{obs:decrease_exploit}
    Ignoring ties, if an action has been continuously selected greedily and is later abandoned, then its Q-value has strictly decreased.
\end{observation}

Consider a transition, triggered by exploration and driven by greedy actions, to another absorbing state. Let action $a$ be the greedy choice. By definition, $Q^a$ exceeds all other Q-values. While playing $a$, only $Q^a$ updates and all other Q-values remain static. If $a$ is eventually abandoned, $Q^a$ must have fallen below one of these fixed values. Therefore, $Q^a$ must have strictly decreased.

This observation is the key to breaking the price cycle. Whenever a bilateral rebound interrupts an alternating-undercut episode, some lower action must have been selected greedily and then abandoned. By Observation~\ref{obs:decrease_exploit}, its Q-value has strictly decreased. The rebound therefore does not return the system to a previous state: \textbf{\textit{it erodes the Q-value bubble}}. Our condition~\ref{order_condition_1} tracks this more fundamental progress in the underlying Q-values beneath the observed price cycle itself.

The need to prioritize Q-values over actions is a key distinction from \cite{dolgopolov2024reinforcement}, who studies the Prisoner's Dilemma with discount factor $\delta=0$. In the Prisoner's Dilemma, the greedy profile $G(s)$ of an absorbing state $s$ is either $(C,C)$ or $(D,D)$. Once the dynamics reach an absorbing state with $G(s)=(D,D)$, no bilateral rebound is possible. Accordingly, the first step in his partial order construction is to compare based on greedy profiles. As discussed above, this approach does not extend to Bertrand competition. Constructing a suitable partial order instead requires prioritizing Q-values over actions. Relatedly, our companion work \citep{Xu2026} extends the analysis to broader competition environments, including logit-demand Bertrand and Cournot environments. There, absorbing states may have asymmetric greedy profiles, requiring a more elaborate analysis that is beyond the scope of this paper.

\subsection{The Double-Edged Role of Exploration}

\begin{figure}[!t]
    \centering
    \subfigure[Random + No]{
    \includegraphics[width=0.49\textwidth]{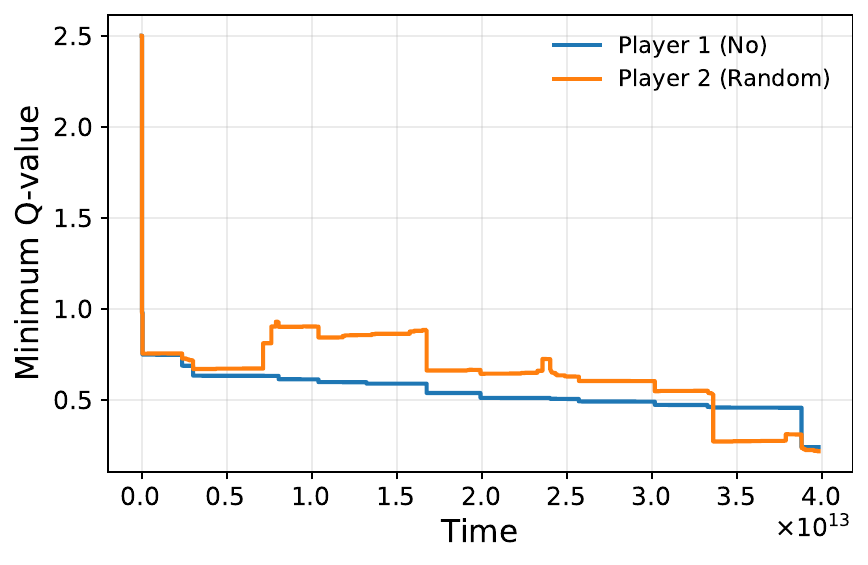}\label{fig:Q_rn}}
    \hfill
    \subfigure[Random + Random]{
    \includegraphics[width=0.49\textwidth]{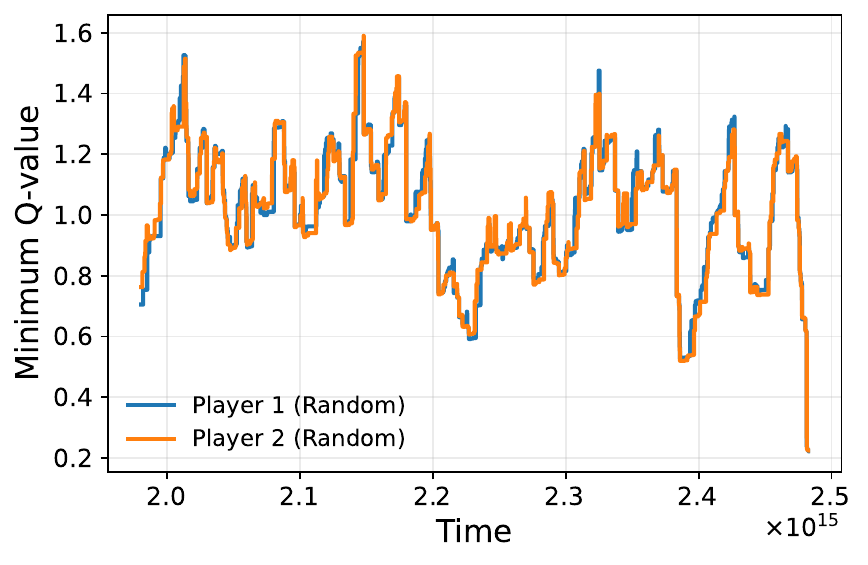}\label{fig:Q_rr}}
    \caption{Q-value dynamics under two exploration profiles. Both panels show 
    a representative sample of the two players' minimal Q-values under the 
    baseline Bertrand simulation of Section~\ref{sec:simulation}, with 
    $\alpha = 0.15$, $\delta = 0.8$, and $\varepsilon = 10^{-10}$. In 
    Panel~(a), only player~$2$ explores (random~$+$~no exploration), and 
    player~$1$'s minimal Q-value declines steadily until the dynamics reach 
    the Nash state. In Panel~(b), both players explore (random~$+$~random), 
    and the Q-value bubble repeatedly re-inflates, delaying convergence.}
\end{figure}

The construction of the order highlights the asymmetric roles of the two players: player~2 carries out the exploration that breaks collusion, while player~1 bears the losses that gradually deflate the Q-value bubble. This asymmetry is the engine of convergence. In our discrete-time setting, one can construct a finite sequence of periods in which only player~2 explores. Along such a path, each cycle of undercutting and rebound further reduces the bubble in player~1's Q-values, pushing them downward step by step. Repeated erosion eventually drives player~1's Q-values for all actions above action~1 below the discounted Nash payoff, so action~1 becomes her unique greedy action regardless of the opponent's choice. Once player~1 is permanently fixed at action~1, action~1 becomes the unique greedy action for player~2 as well, and the dynamics enter the Nash state and remain there. Figure~\ref{fig:Q_rn} traces this process: along the constructed path, player~1's minimal Q-value declines steadily until the dynamics settle at $s^{NE}$.

At first sight, Theorem~\ref{thm:ss} appears hard to reconcile with the simulation evidence in the literature, where supra-competitive outcomes arise under time-decaying exploration, particularly when the discount factor $\delta$ is high (e.g., \citealt{lambin2024less}). The tension arises because actual simulations do not restrict exploration in the way our construction does. Remind that an alternating-undercut episode depresses the Q-values of some lower prices. However, when it is interrupted by a bilateral rebound, exploring those lower prices \textbf{\textit{restores}} their Q-value bubbles and pushes the state back away from $s^{NE}$. 

Therefore, from the perspective of adaptation to NE, the exploration is like a sword with double edges. On the positive side, exploration of profitable actions initiate alternating undercut, driving the state closer to $s^{NE}$; On the negative side, exploration of unprofitable actions inflate and restore Q-values of lower actions, driving the state away from $s^{NE}$.\footnote{\citet{lambin2024less} similarly argue that simultaneous exploration is central to memoryless algorithmic collusion, since a unilateral deviation can break any non-Nash profile but not the NE. This argument alone does not imply convergence, however, as the dynamics could cycle among non-Nash states. Our analysis establishes the crucial step, showing that successive unilateral deviations eventually reach the NE.} In this sense, the dynamics resemble a random walk with $s^{NE}$ as the absorbing boundary. In this random walk, the discount factor governs the scale of both inflation and recovery, since it multiplies the highest Q-value injected in each update. Figure~\ref{fig:Q_rr} illustrates this: when both players explore, the Q-value bubble repeatedly recovers, delaying convergence to the NE; over a long horizon, however, the bubble ultimately disappears and the dynamics reach the NE.

\begin{figure}
    \centering
    \includegraphics[width=\textwidth]{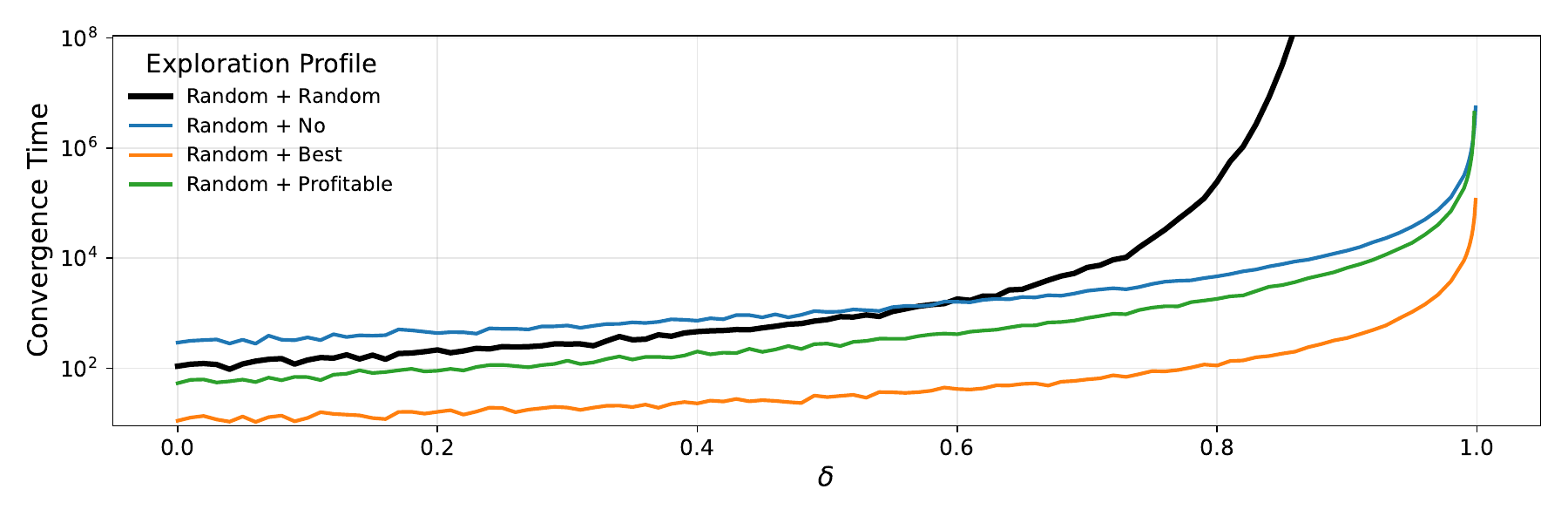}
    \caption{Convergence time to the NE as a function of the discount factor, under various exploration profiles. The simulations use the baseline Bertrand competition of Section~\ref{sec:simulation}, with $\alpha=0.15$ and $\varepsilon=10^{-10}$. The black curve denotes unrestricted $\varepsilon$-greedy exploration; the remaining curves restrict at least one player from exploring unprofitable prices. The exploration profiles are described in Section~\ref{sec:simulation}.}

    \label{fig:various_exploration_mode}
\end{figure}

This double-edged role reconciles our theory with the simulation evidence.  Figure~\ref{fig:various_exploration_mode} reports the expected convergence  time to $s^{NE}$ as a function of the discount factor under several exploration modes. When at least one player is prevented from exploring  unprofitable prices (the colored curves), the dynamics reach $s^{NE}$ in a reasonable time. Even under the unrestricted mode (the black bold curve), the  dynamics still converge to $s^{NE}$, confirming our theory. But the expected convergence time grows explosively in the discount factor, which is why convergence to the NE is rarely observed in simulations, especially when $\delta$ is high.

We conclude this section by two remarks. First, the cycle of symmetric actions, termed as ``spontaneous coupling'' in \cite{banchioartificial}, is the appearance rather than the essence of memoryless algorithmic collusion documented in simulations. Despite the occurrence of such cycles, our order still indicates the state is moving strictly closer to the Nash state. It is the exploration of unprofitable actions during these cycles which recovers Q-values of these actions and therefore slows down the convergence to Nash state in the short run. Second, our theory does not contradict the simulation evidence: \textbf{\textit{the theory describes long-run behavior, whereas the simulations report short-run outcomes}}. The gap between the two is entirely a matter of \textbf{\textit{speed}}. This raises two questions: why does convergence slow so sharply as the discount factor rises, and, more broadly, how do the model's parameters shape the speed of convergence?

\section{Estimation of Convergence Speed}\label{sec:simulation}

This section develops a structural approach to quantify convergence speed under different exploration modes and parameter values. The resulting model allows us to characterize how the learning rate $\alpha$ and the discount factor $\delta$ shape the speed of convergence.

We consider a Bertrand duopoly in which each firm chooses from $100$ prices uniformly spaced between $0.1$ and $1$, selling a homogeneous good valued at $1$. To align the simulations with our theoretical analysis, we fix a constant exploration rate of $\varepsilon=10^{-10}$ and run each experiment until it reaches the NE. Specifically, convergence is defined as the point at which every non-Nash action has a Q-value below the discounted Nash payoff. At this point, no unilateral deviation can move the dynamics away from the NE; conditional on one deviation, a second simultaneous deviation occurs with probability of order $\varepsilon^2$, and is therefore negligible.

We define the \textbf{\textit{(normalized) convergence time}} $T$ as the number of periods until convergence multiplied by $\varepsilon$. Equivalently, convergence occurs after $T/\varepsilon$ periods, so $T$ measures time in units of $1/\varepsilon$, the expected interval between consecutive deviations. This normalization makes $T$ approximately invariant to $\varepsilon$ when $\varepsilon$ is small.\footnote{The invariance follows from two observations. First, simultaneous deviations by both players occur with probability of order $\varepsilon^2$ and are therefore negligible. Second, consecutive single deviations are separated by an expected interval of $1/\varepsilon$ periods, which is sufficiently long for the greedy dynamics to reach an absorbing state between deviations. Hence, the sequence of absorbing states visited between deviations does not depend on $\varepsilon$, and convergence time is determined by the number of deviations required to reach the Nash state, which is captured by $T$. In our simulations, we set $\varepsilon=10^{-10}$; using $\varepsilon=10^{-5}$ produces nearly identical values of $T$.}
For each configuration, we run $100$ independent sessions and report the average convergence time. Appendix~\ref{Appendix:experiment_design} provides additional details on the experimental design.

\subsection{Two Variations of Standard Q-Learning} To recap, the main mechanism underlying random walk nature of the learning dynamics is that, the exploration of unprofitable actions, together with updating with continuation payoff estimation using maximal Q-value, inflate and recover Q-values. Therefore, the exploration affects the learning dynamics through two channels, one on the updating rule the other on chosen actions when exploring. Based on these two channels, we introduce two types of variations to systematically address the role of exploration. First, we introduce a decomposition on updating rule to distinguish the updating part due to exploration. Second, we introduce various exploration modes based on exploring actions. 

\paragraph{Decomposition of Updating Rule.} The standard updating rule can be decomposed as 
\begin{equation}\label{eq:two_learning_rate}
   Q_{i,t+1}^{a} = Q_{i,t}^a + \alpha_1 \left(\frac{u(a_{i,t}, a_{-i,t})}{1-\delta} - Q_{i,t}^a\right) +
   \alpha_2 \left(\max_{a'\in A}Q_{i,t}^{a'} - Q_{i,t}^a \right), 
\end{equation}
where $\alpha_1=\alpha(1-\delta)$ and $\alpha_2=\alpha\delta$. The first term reflects learning from realized payoffs, whereas the second term reflects learning from the continuation value associated with the currently optimal action. If action $a$ is chosen by exploitation, then only the first term appears when updating. Therefore, the second term can be viewed as additional updating part due to exploration. In this sense, such decomposition tries to distinguish the role of exploration in Q-value updating. We thus refer to $\alpha_1$ as the \textbf{\textit{effective learning rate}} and $\alpha_2$ as the \textbf{\textit{explorative learning rate}}. By definition, the original parameter pair $(\alpha,\delta)$ determines both $\alpha_1$ and $\alpha_2$. It is well acknowledged that the discount factor $\delta$ affects the first term through rescaling discounted payoffs and lowering the effective learning rate $\alpha_1$. However, if ignoring its effect on the second term even in Q-learning with single state, like our memoryless setting, then we will overlook its impact on how the updating part driven by exploration changes the Q-value transitions. 

Our estimation strategy is to first examine the role of $(\alpha_1,\alpha_2)$ in the convergence speed and subsequently map the results back to the original parameters $(\alpha,\delta)$.\footnote{Under this decomposition, the discount factor in $u(a_{i,t},a_{-i,t})/(1-\delta)$ only rescales payoffs and therefore does not affect convergence speed. Accordingly, we set $\delta=0$ in the simulation with two learning rates.}

\paragraph{Profiles of Exploration Mode.} To capture the role of exploring actions in learning dynamics, we study four canonical exploration rules. Under \textbf{\textit{Random}} (\texttt{r}) exploration, an exploratory move is drawn uniformly from the action space. Under \textbf{\textit{Best}} (\texttt{b}) exploration, the agent selects the best response. Under \textbf{\textit{Profitable}} (\texttt{p}) exploration, the agent draws uniformly from the set of profitable actions, namely, actions that yield a higher one-stage payoff than the current action given the opponent's action. Under \textbf{\textit{No}} (\texttt{n}) exploration, exploratory moves are not allowed. Combining the exploration rules of the two players yields nine exploration profiles:
\begin{equation}
    \mathcal E= \{\texttt{rr}, \texttt{rb}, \texttt{rp}, \texttt{rn}, \texttt{bb}, \texttt{bp}, \texttt{bn}, \texttt{pp}, \texttt{pn}\}.
\end{equation}


\subsection{Estimation Model}

Due to the random walk nature of learning dynamics, we approximate the complex Markovian process through a simple random-walk representation. We first provide some preliminaries on simple random walk. 
\paragraph{Preliminaries of Simple Random Walk.} In the theoretical analysis, we introduce a partial ordering over states according to their proximity to the Nash state. The learning dynamics can be viewed as moving along this ordering. Following a perturbation, the system either moves closer to the Nash state (a forward step), farther away from it (a backward step), or remains at the same location.\footnote{The theoretical ordering is a partial order consisting of multiple chains that terminate at the Nash state, rather than a single chain. The random-walk representation is therefore a reduced-form approximation that projects the underlying state space onto a one-dimensional distance measure. Because multiple states may be mapped to the same distance level, a perturbation may leave the distance to the Nash state unchanged, generating a stay-put transition. Such transitions also arise in the theoretical order, for example when an exploration occurs at a price above the convergent price.} Motivated by this observation, we approximate the learning process by a random walk.

Specifically, consider a random walk on the finite state space $\{0,1,\ldots,L\}$, where $L$ is the absorbing boundary and $0$ is the reflecting boundary. For $1\le x\le L-1$,
\begin{equation}
X_{t+1}
=
\begin{cases}
X_t+1, & \text{with probability } p,\\
X_t,   & \text{with probability } s,\\
X_t-1, & \text{with probability } q,
\end{cases}
\qquad
p+q+s=1.
\end{equation}
At the reflecting boundary, $P(0\to1)=p$, $P(0\to0)=q+s$.

The state variable \(X_t\) should be interpreted as a latent measure of proximity to the Nash state rather than a literal state count. A forward step corresponds to a perturbation that moves the learning process closer to equilibrium, whereas a backward step corresponds to a perturbation that moves it farther away. The parameter \(L\) therefore measures the effective distance to convergence, while \(p\) and \(q\) summarize the directional tendency.

Under this approximation, the expected hitting time of the absorbing state provides a reduced-form measure of convergence speed. The following proposition gives the expected convergence time starting from the reflecting boundary.

\begin{proposition}\label{prp:hitting_time}
The expected convergence time from state \(0\) is
\begin{equation}\label{eq:hitting_time}
H(L,q,p) =
\begin{cases}
\displaystyle
\frac{L}{p-q}
+
\frac{q}{(p-q)^2}
\left[
\left(\frac{q}{p}\right)^L-1
\right],
&
p\neq q,
\\
\displaystyle
\frac{L(L+1)}{2p},
&
p=q.
\end{cases}
\end{equation}
\end{proposition}

One critical factor on expected convergence time is the sign of the \textbf{\textit{drift}} $\Delta\equiv p-q$, which measures the tendency of the random walk to move toward rather than away from the absorbing boundary. Proposition~\ref{prp:hitting_time} implies that when $\Delta>0$, the expected hitting time grows approximately \textbf{\textit{linearly}} with the length $L$, whereas when $\Delta<0$, it grows approximately \textbf{\textit{exponentially}} with $L$.

The random-walk representation is deliberately parsimonious. Although it abstracts from many features of the underlying Q-learning dynamics, it preserves the key forces governing convergence: the effective distance to equilibrium ($L$) and the directional bias of exploratory moves ($\Delta$). As shown below, these three parameters provide a remarkably accurate description of observed convergence times across learning environments and exploration patterns.

\paragraph{Further Restrictions.}

The simple random walk model can be fully captured by latent parameter profile $(L,p,q)$, which is determined by both effective and explorative learning rate $(\alpha_1,\alpha_2)$ and profile of exploration mode $e$. Our estimation strategy is to first identify the latent parameter profile to minimize residual error. Then, we try to estimate how $(\alpha_1,\alpha_2,e)$ affects the latent parameter profile $(L,p,q)$.

Let $i$ index values of the effective learning rate $\alpha_1$, $j$ index values of the explorative learning rate $\alpha_2$, and $e$ index exploration profiles. Different index not only corresponds to different average convergence time $T_{ije}$, but also different latent random-walk parameter profiles $(L_{ije},p_{ije},q_{ije})$. Since only the average convergence time $T_{ije}$ is observed, these parameters cannot be separately identified without additional restrictions. We therefore impose the following assumptions.

\begin{ass}\label{ass:length}
For each value of $\alpha_1$, the random-walk length $L_i$ depends only on $\alpha_1$ and is decreasing in $\alpha_1$.\footnote{The monotonicity assumption is adopted to facilitate the estimation of latent parameters. Ignoring this assumption does not affect the identification much.}
\end{ass}
 
The length $L_i$ can be interpreted as the minimal number of explorations required to reach the Nash state $s^{NE}$. Therefore, we can at first exclude its dependence on exploration mode profile $e$. Note that the adopted random walk model abstracts away the magnitude of moving forward or backward by taking each step as equal size. To tailor to the estimation model, the minimal number of explorations should be inversely related to the scale of Q-value updating given exploration. In this sense, both $\alpha_1$ and $\alpha_2$ affect length $L$. However, post exploration, Q-values are mainly updated during the phase of alternating undercut without exploration. Therefore, the length $L$ is predominantly determined by the magnitude of updating without exploration, i.e. the effective learning rate $\alpha_1$. 

\begin{ass}\label{ass:prob}
For each pair \((j,e)\), the local transition probabilities \((p_{je},q_{je},s_{je})\) depend only on \(\alpha_2\) and the exploration profile.
\end{ass}

To justify this assumption, remind that the mechanism of the random walk nature is that the exploration of unprofitable actions recovers the Q-value and moves the Q-values away from $s^{NE}$. Therefore, the linkage between exploration mode profile and transition probabilities is obvious. Moreover, as we abstract away the magnitude of moving forward or backward, we link the explorative learning rate $\alpha_2$ with the transition probabilities, to tailor to the adopted simple model.  

Under Assumptions~\ref{ass:length} and~\ref{ass:prob}, convergence times are fully characterized by the length $L_i$ and the transition probabilities $(p_{je},q_{je})$. The model therefore predicts
\[
\widehat T_{ije} = H(L_i,p_{je},q_{je}).
\]

\begin{figure}[!t]
    \centering
    \subfigure[Observed vs. Fitted]{
        \includegraphics[width=0.49\textwidth]{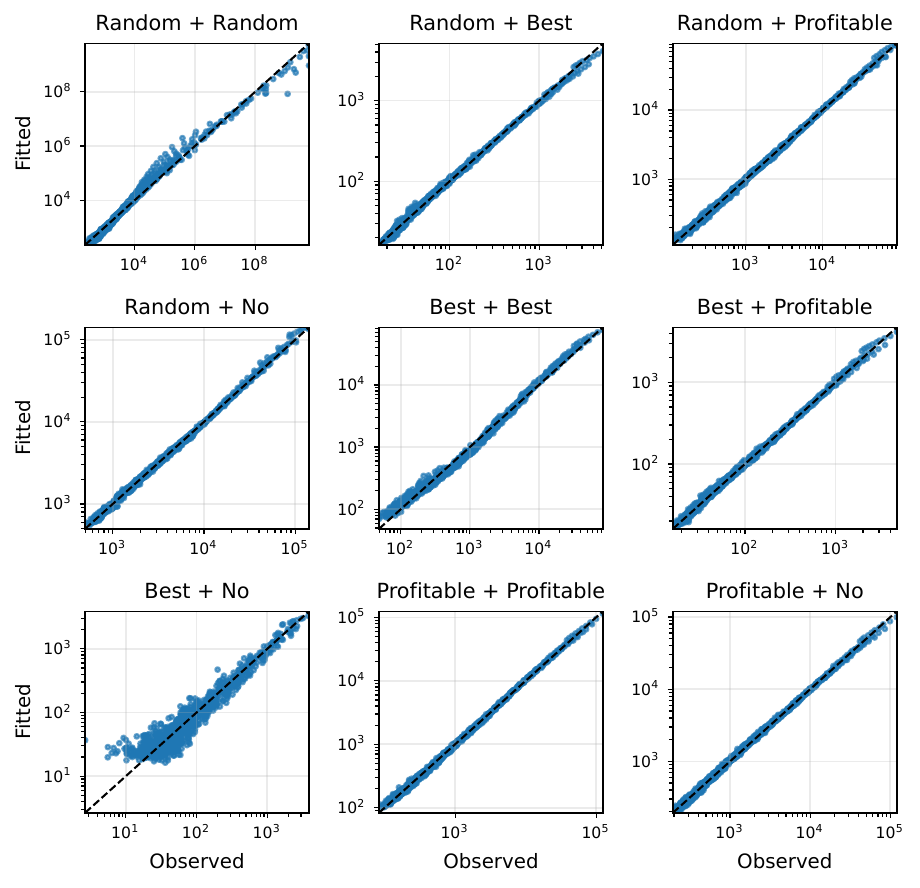}%
        \label{fig:direct_obs_fit}}
    \hfill
    \subfigure[Drift]{
        \includegraphics[width=0.49\textwidth]{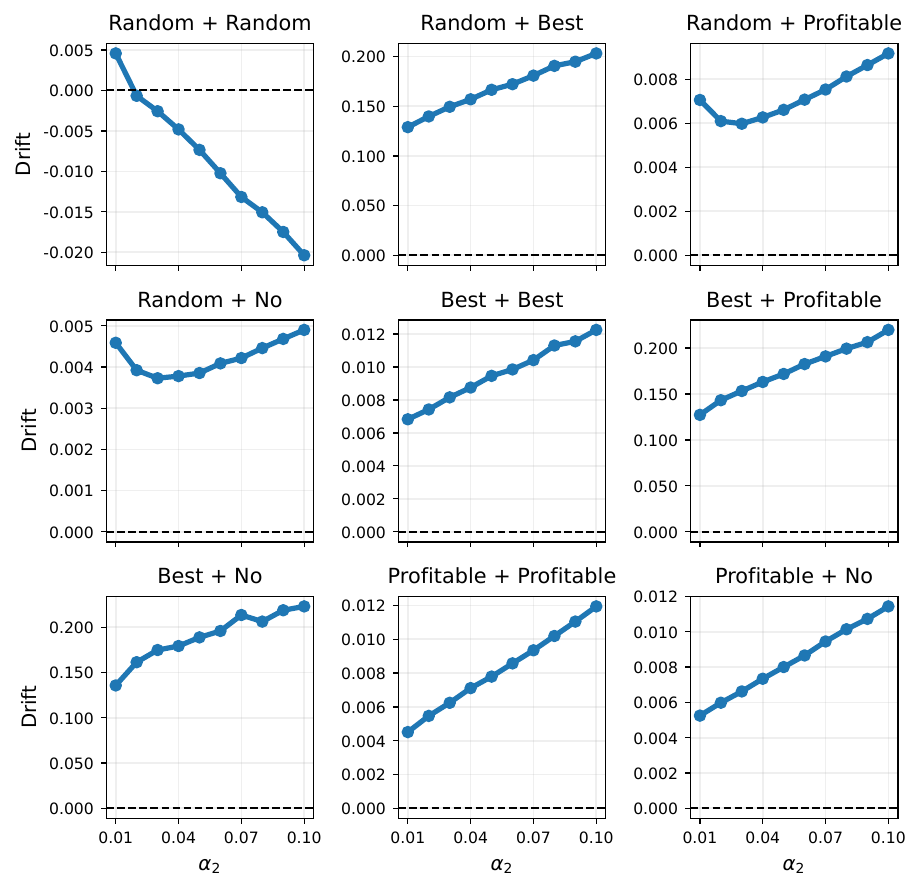}%
        \label{fig:direct_drift}}

    \caption{Fit of the random-walk model. Panel~(a) plots fitted against observed convergence times. Panel~(b) plots the estimated drift  as a function of $\alpha_2$.}
    \label{fig:direct_fit}
\end{figure}

\subsection{Estimation and the Effects of Parameters}

\paragraph{Model Fit.}
To estimate the latent parameters, we form a grid of learning rates, $\alpha_1 \in [0.005, 0.1]$ in steps of $0.001$ and $\alpha_2 \in [0.01, 0.1]$ in steps of $0.01$, and minimize the log-residual between predicted and observed convergence times.\footnote{Some $(\alpha_1,\alpha_2)$ pairs are excluded because their convergence times exceed $10^{10}$ periods and are therefore beyond the range of our simulations. Log-residuals are normalized within each series $(j,e)$ before minimization; see Appendix~\ref{app:estimation} for details.} Figure~\ref{fig:direct_obs_fit} plots observed against fitted convergence times. Most points lie close to the 45-degree line, so the reflecting-boundary random walk captures the main variation in convergence times across $\alpha_1$, $\alpha_2$, and exploration modes.\footnote{The exception is the \texttt{bn} mode, which exhibits large residuals: under \texttt{bn} the dynamics are nearly deterministic, whereas our fixed-step random walk approximates best when enough randomness averages across transitions.}

\paragraph{The Role of the Drift.}
Figure~\ref{fig:direct_drift} plots the estimated drift $\Delta$ against $\alpha_2$ for each exploration profile. A larger $\alpha_2$ amplifies the effect of exploration in both directions. When exploration selects profitable actions, it increases the forward probability; when it selects unprofitable actions, it increases the backward probability by restoring the Q-value inflation of lower actions. The net effect therefore depends on the frequency of unprofitable exploration.

Under \texttt{rr}, both players explore randomly, making unprofitable exploration frequent. As $\alpha_2$ increases, the drift declines steadily and eventually becomes \textbf{\textit{negative}}. This change in the sign of the drift changes the scaling of convergence time from approximately linear in the length $L$ to exponential, making convergence substantially slower. In the other exploration regimes, at least one player is protected from unprofitable exploration. The drift remains positive and often increases with $\alpha_2$. Thus, preventing even one player from selecting unprofitable exploratory actions is sufficient to preserve a positive drift and accelerate convergence.

\begin{figure}[t]
    \centering
    \subfigure[Length and $\alpha_1$]{
    \includegraphics[width=0.49\textwidth]{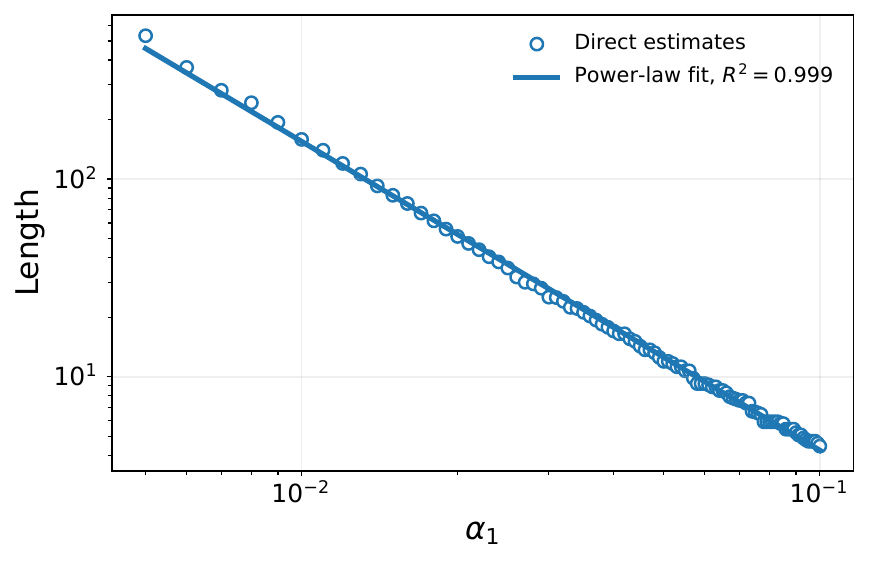}\label{fig:length_alpha1}}
    \hfill
    \subfigure[Transition Probabilities and $\alpha_2$]{\includegraphics[width=0.49\textwidth]{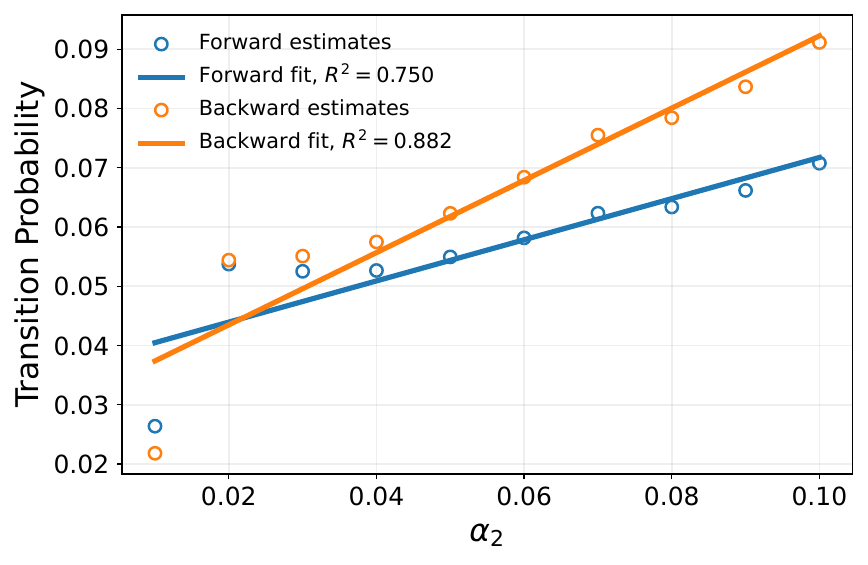}\label{fig:transition_alpha2}}

    \caption{Empirical structure of the length and transition probabilities. Panel~(a) plots the estimated length against $\alpha_1$ on a log-log scale. Panel~(b) plots the estimated forward and backward transition probabilities under the \texttt{rr} mode against $\alpha_2$.}
\label{fig:structure}
\end{figure}

\paragraph{Structures of Latent Parameters.}
We now map these estimates to the primitive parameters, the learning rate $\alpha = \alpha_1 + \alpha_2$ and the discount factor $\delta = \alpha_2/(\alpha_1 + \alpha_2)$, focusing on the \texttt{rr} mode. Because the estimated grid covers only a limited range of $(\alpha, \delta)$, we exploit two empirical regularities in Figure~\ref{fig:structure}: the length $L$ is nearly log-log linear in $\alpha_1$, and, under \texttt{rr}, the transition probabilities are approximately linear in $\alpha_2$. Imposing these two functional forms lets us predict convergence times for $(\alpha, \delta)$ outside the estimation grid and test the predictions against fresh simulations.

\begin{figure}[t]
    \centering
    \subfigure[Discount Factor]{\includegraphics[width=0.49\textwidth]{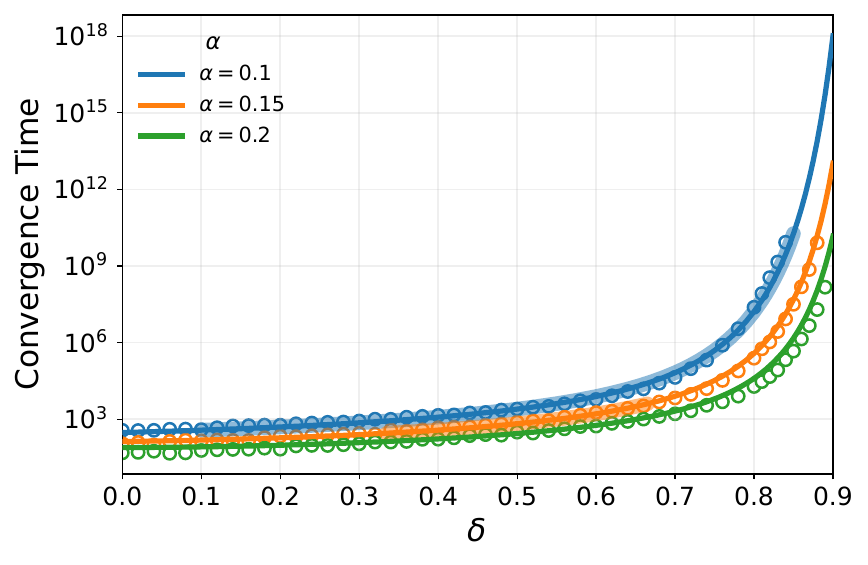}\label{fig:delta_effect}}
    \hfill
    \subfigure[Learning Rate]{\includegraphics[width=0.49\textwidth]{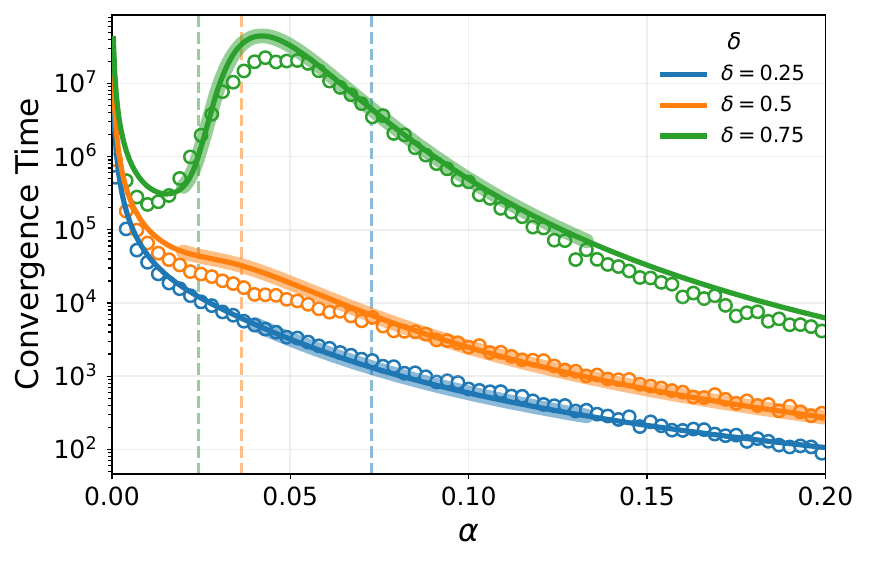}\label{fig:alpha_effect}}

    \caption{Predicted and simulated convergence times under the structured random-walk model. Points denote simulation results and the solid line denotes model predictions. The shaded region corresponds to parameter values included in the estimation sample; the unshaded region represents out-of-sample extrapolation. In Panel (b) the dashed line represents the point where $\Delta = 0$.}
    \label{fig:delta_alpha_effect}
\end{figure}

\paragraph{Effect of the Discount Factor.}
Fixing the learning rate $\alpha$, increasing the discount factor $\delta$ substantially increases the convergence time $T$. A higher $\delta$ lowers the effective learning rate $\alpha_1$, increasing the length $L$, and raises the explorative learning rate $\alpha_2$, reducing the drift $\Delta$. The random walk therefore becomes both longer and less directed toward convergence. Figure~\ref{fig:delta_effect} shows that the resulting predictions closely track the simulation results, both within and outside the estimation region, suggesting that these two channels capture the primary effect of the discount factor on convergence.

\paragraph{Effect of the Learning Rate.}
The effect of the learning rate $\alpha$ is more subtle and can be \textbf{\textit{non-monotonic}}. Lowering $\alpha$ creates two opposing forces: it reduces the effective learning rate $\alpha_1$, which increases the length $L$ and raises the convergence time $T$; but it also reduces the explorative learning rate $\alpha_2$, which strengthens the drift $\Delta$ toward the NE and lowers $T$. Which force dominates depends on the sign of $\Delta$: when $\Delta < 0$, $T$ grows exponentially in $L$, whereas when $\Delta > 0$, it grows only linearly. As $\alpha$ falls and $\Delta$ turns positive, $T$ therefore drops from its exponential to its linear level, and this drop is larger when $L$ is larger. 

The magnitude of the drop depends on the discount factor $\delta$. Since the sign change occurs at a fixed threshold $\alpha_2 = \delta\alpha$ (dashed line in Figure~\ref{fig:alpha_effect}), a smaller $\delta$ places it at a larger $\alpha$, where $\alpha_1$ is large and $L$ is small. The drop is then too small to outweigh the rise from the longer walk, so $T$ rises monotonically as $\alpha$ falls. A larger $\delta$ places the threshold at a smaller $\alpha$, where $L$ is large and the drop is substantial. However, as $\alpha$ falls further, $L$ keeps growing while $\Delta$ approaches a positive limit, so $T$ rises again. Figure~\ref{fig:alpha_effect} confirms this pattern: at $\delta = 0.75$, $T$ rises, then falls sharply, and finally rises again as $\alpha$ decreases, whereas at $\delta = 0.5$ and $0.25$ it remains monotone.

\section{Application}\label{sec:application}

\subsection{Prisoner's Dilemma}\label{subsec:pd}

In this subsection, we focus on the Prisoner's Dilemma (PD), the simplest environment for studying algorithmic collusion. Figure~\ref{fig:pd_game} presents the payoff matrices. Panel (a) shows a general PD, and Panel (b) shows the specification in \cite{banchioartificial}, both in normalized form. Action $C$ denotes cooperation and action $D$ denotes defection.

\begin{figure}[htbp]
\centering

\subfigure[General Setting: $u_{DC}>1>u_{DD}>0$]{
\begin{tabular}{c c|c|c|}
\multicolumn{2}{c}{} & \multicolumn{2}{c}{Player 2} \\
\multicolumn{2}{c}{} & \multicolumn{1}{c}{$C$} & \multicolumn{1}{c}{$D$} \\ \cline{3-4}
\multirow{2}{*}{Player 1}
& $C$ & $1,\,1$ & $0,\,u_{DC}$ \\ \cline{3-4}
& $D$ & $u_{DC},\,0$ & $u_{DD},\,u_{DD}$ \\ \cline{3-4}
\end{tabular}}
\hfill
\subfigure[One-parameter Specification: $1<g<2$]{
\begin{tabular}{c c|c|c|}
\multicolumn{2}{c}{} & \multicolumn{2}{c}{Player 2} \\
\multicolumn{2}{c}{} & \multicolumn{1}{c}{$C$} & \multicolumn{1}{c}{$D$} \\ \cline{3-4}
\multirow{2}{*}{Player 1} 
 & $C$ & $1,\,1$ & $0,\,\nicefrac{2}{g}$ \\ \cline{3-4}
& $D$ & $\nicefrac{2}{g},\,0$ & $\nicefrac{2}{g}-1,\,\nicefrac{2}{g}-1$ \\ \cline{3-4}
\end{tabular}}

\caption{Payoffs of the Prisoner's Dilemma. Panel (a) presents a general specification. Panel (b) shows the specific setting in \cite{banchioartificial}.}
\label{fig:pd_game}
\end{figure}

\begin{figure}[t]
    \centering
    \subfigure[Exploration Rate]{
    \includegraphics[width=0.325\textwidth]{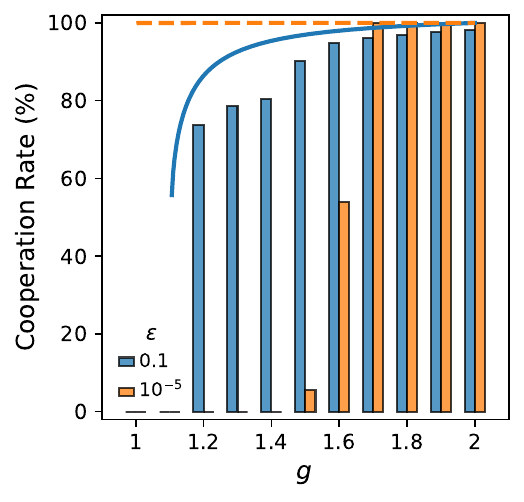}\label{fig:banchio_exploration}}
    \hfill
    \subfigure[Discount Factor]{
    \includegraphics[width=0.325\textwidth]{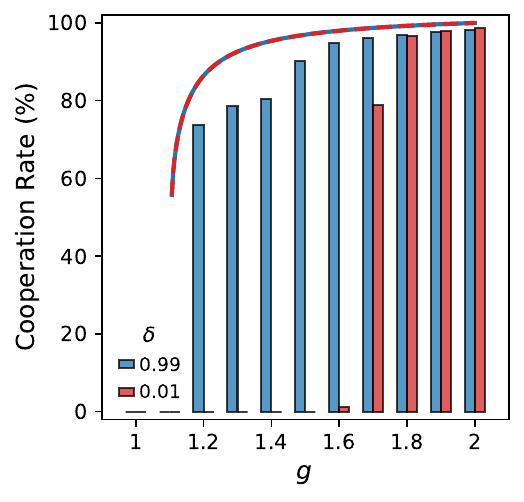}\label{fig:banchio_discount}}
    \hfill
    \subfigure[Learning Rate]{\includegraphics[width=0.325\textwidth]{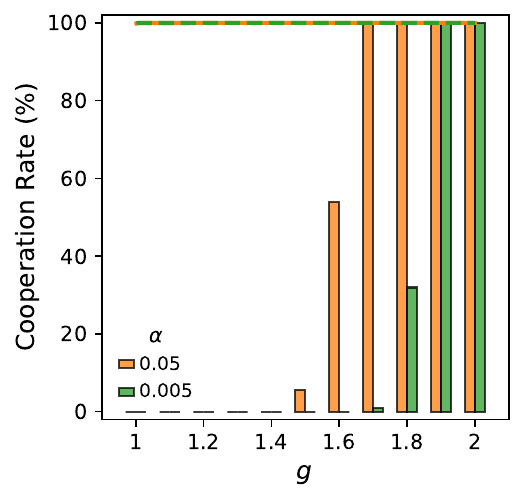}\label{fig:banchio_alpha}}

    \caption{
    Cooperation rate in the Prisoner's Dilemma. Each bar reports the fraction of periods in the final 20\% of a simulation in which cooperation is the player's greedy action. For each parameter profile $(\alpha,\varepsilon)$, the simulation horizon is set to $500/(\alpha\varepsilon)$ periods, corresponding to 100,000 periods under the benchmark specification $(\alpha,\varepsilon)=(0.05,0.1)$ in \cite{banchioartificial}. Lines report the cooperation rates predicted by the theory of \cite{banchioartificial}. Panel (a) fixes $\alpha=0.05$ and $\delta=0.99$ while varying $\varepsilon\in\{0.1,10^{-5}\}$. Panel (b) fixes $\alpha=0.05$ and $\varepsilon=0.1$ while varying $\delta\in\{0.99,0.01\}$. Panel (c) fixes $\delta=0.99$ and $\varepsilon=10^{-5}$ while varying $\alpha\in\{0.05,0.005\}$.
    }
    \label{fig:banchio}
\end{figure}

\cite{banchioartificial} is a seminal contribution to explaining the emergence of memoryless algorithmic collusion. Using a continuous-time approximation of the learning dynamics, the authors show that when the exploration rate $\varepsilon$ is small, there exists a region of Q-value initializations from which play converges to a cycle alternating between $CC$ and $DD$. They also characterize the long-run fraction of time spent in the cooperative state $CC$. The continuous-time approximation, however, implicitly requires the learning rate to vanish. When the learning rate is bounded away from zero, their theoretical predictions can diverge substantially from simulated behavior. Although \cite{banchioartificial} find that their theory matches simulations well under their baseline parameters, Figure~\ref{fig:banchio} shows that the match is weaker under other parameter values.

First, consider the exploration rate (Figure~\ref{fig:banchio_exploration}). Their theory predicts that lower exploration increases the long-run frequency of cooperation and widens the range of $g$ over which cooperative cycles are sustained. The simulations show the opposite pattern for moderate $g$. Reducing the exploration rate from $0.1$ to $10^{-5}$ lowers the cooperation rate for $g \leq 1.6$, and cooperation emerges only for $g \geq 1.6$ rather than near $g \approx 1$. Second, consider the discount factor (Figure~\ref{fig:banchio_discount}). The theory predicts that the cooperation rate is independent of discounting. In the simulations, reducing the discount factor from $0.99$ to $0.01$ substantially lowers the cooperation rate for $g \leq 1.6$. Third, consider the learning rate, holding the exploration rate fixed at $\varepsilon = 10^{-5}$ (Figure~\ref{fig:banchio_alpha}). If the approximation were accurate, a smaller learning rate should bring the simulations closer to the theory. Instead, cooperation becomes harder to sustain. When the learning rate falls from $\alpha = 0.05$ to $\alpha = 0.005$, cooperation emerges only for substantially larger values of $g$, approximately $g \geq 1.8$.

Taken together, these results suggest that the discrepancies arise from the finite-step-size nature of Q-learning. The continuous-time approximation captures the limiting behavior as $\alpha \to 0$, but it can misrepresent the dynamics when the learning rate is bounded away from zero. Under the continuous-time approximation, Q-values are updated by their expected increments, averaged over the deviation probabilities; starting from symmetric Q-values, the two players therefore remain symmetric throughout. Under discrete-time updating, by contrast, Q-values move by their realized increments, and this symmetry breaks down. Our result relies precisely on this asymmetry: although near-symmetry may persist for a long time, once the asymmetry accumulates and the dynamics fall into the NE, escape becomes difficult. Our theory thus complements the continuous-time approach: we allow a strictly positive learning rate and instead take the exploration rate to zero.

\begin{figure}[t]
    \centering
    \subfigure[Payoffs]{\includegraphics[width=0.325\textwidth]{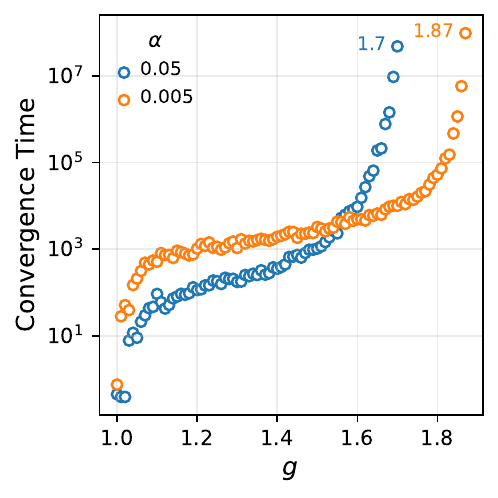}\label{fig:pd_payoff}}
    \hfill
    \subfigure[Discount Factor]{
    \includegraphics[width=0.325\textwidth]{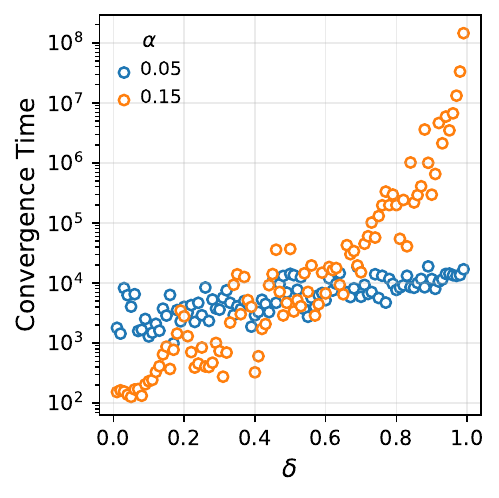}\label{fig:pd_delta}}
    \hfill
    \subfigure[Learning Rate]{
    \includegraphics[width=0.325\textwidth]{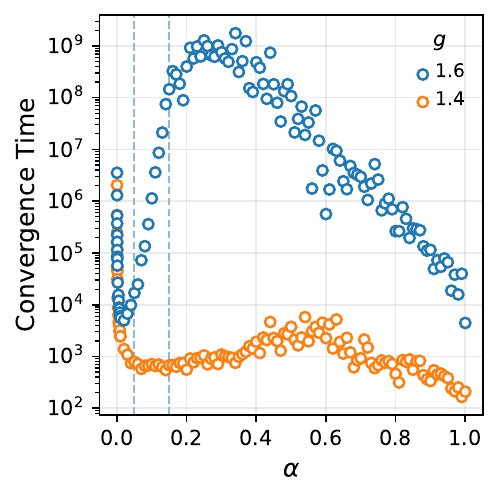}\label{fig:pd_alpha}}
    
    \caption{Convergence times in the Prisoner's Dilemma. In all panels, \(\varepsilon = 10^{-5}\). Panel (a) fixes \(\delta=0.99\) and considers \(\alpha\in\{0.05,0.005\}\).  Panel (b) fixes \(g=1.6\) and considers \(\alpha\in\{0.05, 0.15\}\). Panel (c) fixes \(\delta=0.99\) and considers \(g\in\{1.4,1.6\}\).}
    \label{fig:pd_convergence}
\end{figure}

When the exploration rate is sufficiently low, the dynamics converge to the NE for a wide range of payoffs within our computational budget (Figure~\ref{fig:pd_payoff}), consistent with the theoretical predictions. In particular, when $\alpha=0.005$, play eventually converges to mutual defection even at $g=1.87$, where $u_{DD}\approx0.07$ and $u_{DC}\approx1.07$, so that defection yields only a 7\% payoff advantage.

Figure~\ref{fig:pd_payoff} also shows that the convergence time increases sharply with $g$. At first glance, this pattern appears inconsistent with the Bertrand results, since defection is always a profitable deviation in the PD. The key, however, is that profitability is a matter of degree. As $g$ increases, the gain from defection shrinks. When this gain becomes sufficiently small, a profitable deviation behaves much like an unprofitable one from the perspective of Q-learning, and the dynamics increasingly resemble the \texttt{rr} regime identified in Bertrand competition.\footnote{In Bertrand competition with finely discretized prices, a marginal undercut nearly doubles the deviator's profit relative to collusion, corresponding to approximately $g\approx1$ in the PD. Discretizing the price space means that prices can no longer be adjusted infinitesimally. By contrast, in the PD, a large value of $g$ makes defection only marginally more profitable than cooperation.} Consequently, the drift toward the Nash state declines with $g$, leading to substantially slower convergence.

The effects of the discount factor and learning rate follow the same mechanism as in Bertrand competition. Figure~\ref{fig:pd_alpha} shows that when $g=1.4$, deviations remain sufficiently profitable that the drift becomes positive at relatively high learning rates, and a higher learning rate uniformly accelerates convergence. By contrast, when $g=1.6$, the drift turns positive only after the length has become large, generating the same non-monotonic pattern observed in Bertrand competition. Figure~\ref{fig:pd_delta} shows a similar role for the discount factor when $g=1.6$. For $\alpha=0.05$ (the first dashed line in Figure~\ref{fig:pd_alpha}), the drift remains positive even when $\delta=0.99$. Increasing the discount factor therefore mainly increases the length of random walk, leading to only a modest increase in convergence time. For $\alpha=0.15$ (the second dashed line), increasing $\delta$ also turns the drift negative, leading to a sharp increase in convergence time. Overall, the length and drift decomposition developed for Bertrand competition continues to account for the convergence patterns observed in the PD.

\subsection{Auction}

Auctions are among the most prominent environments in which algorithms are deployed to bid. The two basic formats are the first-price auction (FPA) and the second-price auction (SPA). \cite{banchio2022artificial} find that Q-learning bidders converge to the NE under the SPA but sustain supra-competitive outcomes under the FPA. Our theory clarifies this contrast: in the long run, both formats converge to the same NE, but the auction format affects the speed of convergence.

\begin{figure}[t]
    \centering
    \subfigure[Weight]{\includegraphics[width=0.325\textwidth]{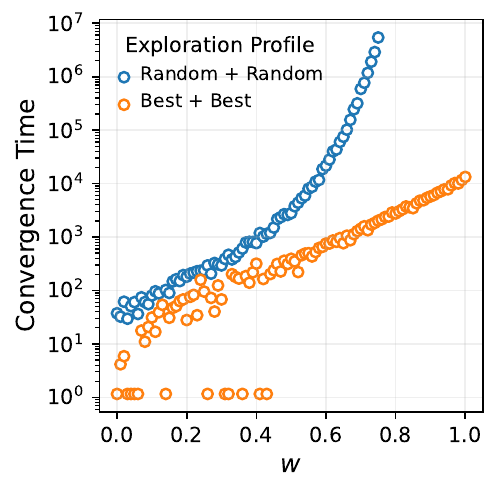}\label{fig:auction_weight}}
    \hfill
    \subfigure[Discount Factor]{\includegraphics[width=0.325\textwidth]{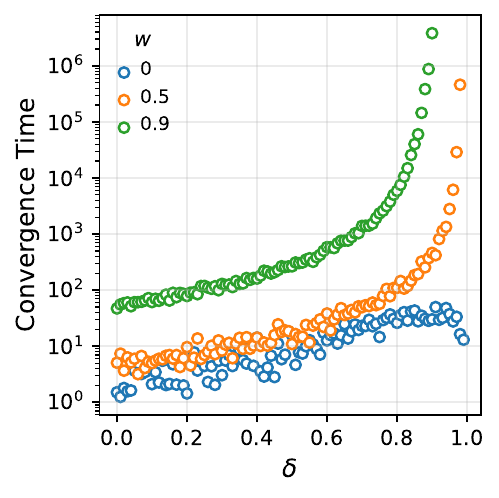}\label{fig:auction_delta}}
    \hfill
    \subfigure[Learning Rate]{\includegraphics[width=0.325\textwidth]{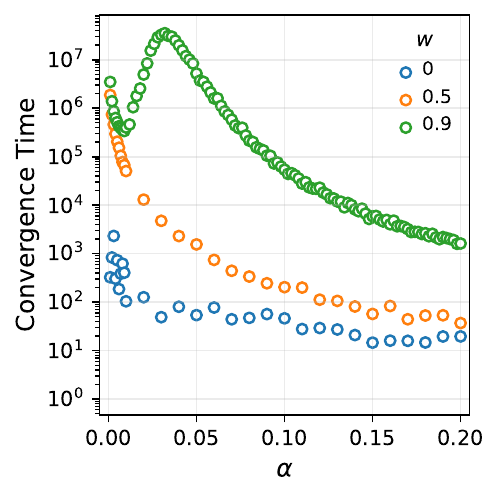}\label{fig:auction_alpha}}
    
    \caption{Convergence times in the auction game. In all panels, \(\varepsilon = 10^{-10}\). Panel (a) fixes \(\delta = 0.95\) and \(\alpha = 0.15\), and also reports convergence under the best+best exploration mode. Panel (b) fixes \(\alpha = 0.15\). Panel (c) fixes \(\delta = 0.8\). Panels (b) and (c) consider \(w \in \{0, 0.5, 0.9\}\).}
    \label{fig:auction_convergence}
\end{figure}

To move between the two formats continuously, let $w$ denote the weight placed on the first-price rule: the winner with bid $b$ pays $w b + (1 - w) b'$, where $b'$ is the opponent's bid. Figure~\ref{fig:auction_weight} shows how $w$ affects the convergence time. The weight plays the same role as $g$ in the PD above. Even in the \texttt{bb} exploration profile, the convergence time increases with the weight: its log is roughly linear in $w$, rising from a single exploration to over $10^4$ explorations. This explains why the FPA in \cite{banchio2022artificial} sustains short-run collusion, even under local deviations.  But the situation with nearby bids is not yet severe. Once bidders can explore higher bids, the log convergence time increases exponentially in the weight. The contrast with \cite{banchio2022artificial} is instructive. They attribute the failure of the FPA to converge to bidders' incentive to deviate to nearby bids. Our analysis suggests otherwise: it is the deviations toward higher bids that make convergence under the FPA prohibitively slow.

Figure~\ref{fig:auction_delta} and \ref{fig:auction_alpha} plots the effects of the discount factor and learning rate on the convergence time for different weights. The results parallel those in the PD. When the weight is small, so that deviations are profitable, the discount factor has little effect on the convergence speed and the learning rate has a monotone effect. When the weight is large, the discount factor substantially slows down convergence and the learning rate has a non-monotonic effect.

\subsection{Sequential Entry}

Consider a sequential entry game: an incumbent initially operates alone in the market, and an entrant arrives after some time. When both firms price with Q-learning, the entrant, being new to the environment, explores intensively, while the incumbent has largely finished exploring. \cite{lambin2024less} study an extreme version of this asymmetry: the incumbent explores only before entry, and after entry only the entrant explores. They find that play then converges to the NE. This setting corresponds exactly to our \texttt{rn} exploration profile. When the incumbent stops exploring, its Q-values decline steadily, so it settles on the Nash price, which in turn leads the entrant to the Nash price as well.

\subsection{Synchronous Updating}

Our main analysis focuses on asynchronous updating: in each period, only the chosen action's Q-value is updated, and the others remain unchanged. The alternative is synchronous updating, under which the agent updates all actions through counterfactual estimation. The simplest form uses the payoff of each action against the opponent's realized action:
\begin{equation}\label{eq:synchronous}
Q^{a}_i(s_{t+1}) = (1-\alpha) Q^{a}_i(s_t) + \alpha \left[u(a, a_{-i}) + \delta \max_{a'} Q_i^{a'} (s_t)\right], \quad \forall a \in A.
\end{equation}

\begin{proposition}\label{prop:syn}
    In the class of Bertrand-style competition games, standard Q-learning with synchronous updating \eqref{eq:synchronous} has $\{s^{NE}\}$ as its unique 
    recurrent class.
\end{proposition}

While \cite{asker2023impact} establishes that $s^{NE}$ is the unique absorbing state (termed as rest point in their paper), we extend this result by proving that it is also the unique recurrent state. \cite{asker2023impact} finds that Q-learning agents with synchronous updating converge to the NE in Bertrand competition. This finding is easy to explain within our framework. Under synchronous updating, the NE is not only stochastically stable but also the unique recurrent state: unless each agent's action is a best response to the other's, the Q-value of the best response must eventually exceed that of the current action. Hence, from any initialization, the dynamics converge to the NE without any exploration, and questions of stochastic stability and convergence speed become moot.

\cite{decarolis2023artificial} also compares asynchronous and synchronous updating, in a SPA environment with three asymmetric bidders. In contrast to \cite{asker2023impact}, they find that synchronous updating reduces competition among bidders. The key difference is that their game has multiple equilibria. Our framework does not speak to this case, as we focus on games with a unique equilibrium. Which equilibrium the dynamics select under asynchronous versus synchronous updating is an open question that we leave for future work.

\newpage

\bibliographystyle{chicago}
\bibliography{refs}

\newpage

\appendix

\section{The Discrete Updating Rule}
\label{appendix:discrete_updating}

\subsection{The Fixed Point of the Frozen Dynamics}

At any state $s_t$, suppose the profile $(a_i,a_{-i})$ is played in every period. This profile need not be greedy. By \eqref{upd:async}, in each such period only $Q_i^{a_i}$ is revised, while every other value $Q_i^{a'}$, $a'\neq a_i$, stays fixed at its level $Q_i^{a'}(s_t)$. We call the resulting evolution of the single value $q:=Q_i^{a_i}$ the \textbf{\textit{frozen dynamics}}: the realized payoff $u(a_i,a_{-i})$ and the other actions' values $\max_{a'\neq a_i}Q_i^{a'}(s_t)$ are held fixed, and only $q$ is updated. Under the frozen dynamics the target \eqref{eq:target} reads
\[
    V_i(a_i,a_{-i};s_t)
    =u(a_i,a_{-i})+\delta\max\Bigl\{\,q,\ 
        \max_{a'\neq a_i}Q_i^{a'}(s_t)\,\Bigr\},
\]
and we seek its fixed point, that is, the value of $q$ to which the repeated updating of $Q_i^{a_i}$ under $(a_i,a_{-i})$ would converge.

\begin{lemma}[Fixed point]\label{lem:fp}
Viewed as a map of $q$, $q\mapsto u(a_i,a_{-i}) +\delta\max\{q,\,\max_{a'\neq a_i}Q_i^{a'}(s_t)\}$ is a contraction with modulus $\delta<1$, and has the unique fixed point
\begin{equation}\label{eq:fixed_point}
    Q^\ast_i(a_i,a_{-i};s_t)
    =\max\Bigl\{\frac{u(a_i,a_{-i})}{1-\delta},\
    u(a_i,a_{-i})+\delta\max_{a'\neq a_i}Q_i^{a'}(s_t)\Bigr\}.
\end{equation}
Moreover,
\begin{equation}\label{eq:pull}
    V_i(a_i,a_{-i};s_t)-Q^\ast_i(a_i,a_{-i};s_t)
    =\theta\,\bigl(Q_i^{a_i}(s_t)-Q^\ast_i(a_i,a_{-i};s_t)\bigr),
    \qquad \theta\in [0,\delta],
\end{equation}
so the target lies on the same side of $Q^\ast_i(a_i,a_{-i};s_t)$ as $Q_i^{a_i}(s_t)$ and $|V_i(a_i,a_{-i};s_t)-Q^\ast_i(a_i,a_{-i};s_t)| \le\delta\,|Q_i^{a_i}(s_t)-Q^\ast_i(a_i,a_{-i};s_t)|$: the target always points toward the fixed point.
\end{lemma}

\begin{proof}
Write $m:=\max_{a'\neq a_i}Q_i^{a'}(s_t)$, $u:=u(a_i,a_{-i})$, $V(q):=u+\delta\max\{q,m\}$, and $Q^\ast:=Q^\ast_i(a_i,a_{-i};s_t)$.

Since $x\mapsto\max\{x,m\}$ is $1$-Lipschitz, $|V(q)-V(q')|\le\delta|q-q'|$. As $\delta<1$, $V$ is a contraction and thus admits a unique fixed point $Q^\ast$ by the Banach fixed point theorem. The fixed point satisfies $q=u+\delta\max\{q,m\}$. If $q\ge m$, then $q=u/(1-\delta)$, feasible iff $u\ge(1-\delta)m$; if $q<m$, then $q=u+\delta m$, feasible iff $u<(1-\delta)m$. Since these cases are mutually exclusive and exhaustive, $Q^\ast=\max\{u/(1-\delta),\,u+\delta m\}$.

Since $Q^\ast$ is a fixed point, $V(q)-Q^\ast=V(q)-V(Q^\ast)=\delta(\max\{q,m\}-\max\{Q^\ast,m\})$. For $q\neq Q^\ast$, define $\theta:=\delta\frac{\max\{q,m\}-\max\{Q^\ast,m\}}{q-Q^\ast}$. Then $V(q)-Q^\ast=\theta(q-Q^\ast)$. Since $x\mapsto\max\{x,m\}$ is nondecreasing and $1$-Lipschitz, $0\le\frac{\max\{q,m\}-\max\{Q^\ast,m\}}{q-Q^\ast}\le1$, and therefore $0\le\theta\le\delta$. For $q=Q^\ast$, the identity is trivial. Hence $(V(q)-Q^\ast)(q-Q^\ast)\ge0$ and $|V(q)-Q^\ast|\le\delta|q-Q^\ast|$.
\end{proof}

\subsection{Admissible Updates}

Let $\mathcal{D}$ be the grid, with $\eta$ the largest gap between adjacent points. Under discrete updating, an update move the played action's value toward its \emph{current} fixed point. Fix a constant $C \geq 1$. When $a_i$ is played against $a_{-i}$ at $s_t$, write $Q^\ast := Q^\ast_i(a_i, a_{-i}; s_t)$ and $q := Q_i^{a_i}(s_t)$; the update is a deterministic map $q \mapsto q^+ \in \mathcal{D}$ satisfying the three conditions below.

\begin{ass}[Contraction outside the band]\label{ax:contract}
    If $|q - Q^\ast| > C\eta$, then
    \[
        |q^+ - Q^\ast| < |q - Q^\ast|
        \quad\text{and}\quad
        \Bigl[(q^+ - Q^\ast)(q - Q^\ast) \geq 0
            \ \text{or}\ |q^+ - Q^\ast| \leq C\eta\Bigr].
    \]
\end{ass}

Assumption~\ref{ax:contract} is well defined on the grid: when $|q - Q^\ast| > C\eta \geq \eta$, the gap between $q$ and $Q^\ast$ exceeds one grid spacing, so some grid point strictly between them exists and satisfies the condition. It is the discrete counterpart, in the outer region, of \eqref{upd:closer} and \eqref{upd:no-overshoot}, and is in fact more general. The main-text rule is stated relative to the target $V_i(a_i, a_{-i}; s_t)$, whereas Assumption~\ref{ax:contract} is stated relative to the fixed point $Q^\ast$; by \eqref{eq:pull}, $V_i - Q^\ast = \theta\,(q - Q^\ast)$ with $\theta \in [0, \delta]$, so the target lies on the same side of $Q^\ast$ as $q$. Moving strictly toward $V_i$ without overshooting therefore also moves strictly toward $Q^\ast$ without overshooting, so any update satisfying \eqref{upd:closer}--\eqref{upd:no-overshoot} satisfies Assumption~\ref{ax:contract}.

\begin{ass}[Absorption in the band]\label{ax:absorb}
    If $|q - Q^\ast| \leq C\eta$, then $|q^+ - Q^\ast| \leq C\eta$.
\end{ass}

\begin{lemma}[Convergence to the band]\label{lemma:band-convergence}
    Suppose $(a_i, a_{-i})$ is played over a segment during which $Q^\ast_i(a_i, a_{-i}; s_t)$ stays constant, denoted $Q^\ast$. Under Assumptions~\ref{ax:contract} and~\ref{ax:absorb}, $Q_i^{a_i}$ enters the band $\{q \in \mathcal{D} : |q - Q^\ast| \leq C\eta\}$ after finitely many plays and remains there.
\end{lemma}

\begin{proof}
    While $|Q_i^{a_i} - Q^\ast| > C\eta$, Assumption~\ref{ax:contract} strictly reduces the distance to $Q^\ast$. Since only finitely many grid points lie between $Q_i^{a_i}$ and $Q^\ast$, the band is reached in finitely many plays, and Assumption~\ref{ax:absorb} keeps $Q_i^{a_i}$ within it thereafter.
\end{proof}

Since the update is deterministic and the $C\eta$-band is finite, iterating it from a value $q$ eventually cycles through a finite set of grid points, called the limit set $R(q)$.

\begin{ass}[Monotone entry]\label{ax:monotone}
    For any two values $q \geq \tilde q$ updated toward the same fixed point $Q^\ast$, the limit sets they reach satisfy $\min R(q) \geq \min R(\tilde q)$.
\end{ass}

Assumption~\ref{ax:monotone} ensures that if a perturbation briefly raises a greedy action's Q-value above the band, the action remains greedy once its Q-value re-enters. Under Assumptions~\ref{ax:monotone}, the limit sets form an interval partition of the band: every value starting outside the band reaches the outermost limit set, while each inner limit set is reached only from initial values already within it.

\begin{definition}[Admissible update]\label{def:admissible}
A state transition $s_t \mapsto s_{t+1}$ is \textbf{\textit{admissible}} if the played action's Q-value is updated to $Q_i^{a_i}(s_{t+1}) = q^+$ according to Assumptions~\ref{ax:contract}--\ref{ax:monotone}. The transition is \textbf{\textit{asynchronous}} if the unplayed actions are not updated, i.e., $Q_i^a(s_{t+1}) = Q_i^a(s_t)$ for all $a \neq a_i$, and \textbf{\textit{synchronous}} if their Q-values are also updated according to Assumptions~\ref{ax:contract}--\ref{ax:monotone}, each toward its own fixed point $Q^\ast_i(a, a_{-i}; s_t)$.
\end{definition}

\subsection{Standard Q-learning is Admissible}

Standard Q-learning \eqref{eq:standardQ} with rate $\alpha_i \in (0,1]$ forms the off-grid value $(1-\alpha_i) Q_i^{a_i}(s_t) + \alpha_i V_i(a_i,a_{-i};s_t)$ and rounds it to the nearest grid point in $\mathcal{D}$, an adjustment of at most $\eta/2$. Throughout, write $Q^\ast := Q^\ast_i(a_i, a_{-i}; s_t)$.

\begin{claim}[Q-learning is admissible]\label{prop:qadmissible}
    Standard Q-learning \eqref{eq:standardQ} with rounding to the nearest grid point satisfies Assumptions~\ref{ax:contract}--\ref{ax:monotone} with
    \[
        C \geq \max\Bigl\{\, 1,\ 
        \max_{i \in I} \frac{1}{2\,\alpha_i(1-\delta)} \,\Bigr\}.
    \]
\end{claim}

\begin{proof}
Fix a player $i$ and write $V_i := V_i(a_i, a_{-i}; s_t)$, $d := |Q_i^{a_i}(s_t) - Q^\ast|$, and $\lambda := 1 - \alpha_i(1-\delta)$. By \eqref{eq:pull}, $V_i - Q^\ast = \theta\,(Q_i^{a_i}(s_t) - Q^\ast)$ for some $\theta \in [0, \delta]$, so the pre-rounding value satisfies
\[
    (1-\alpha_i) Q_i^{a_i}(s_t) + \alpha_i V_i - Q^\ast
    = \bigl[(1-\alpha_i) + \alpha_i \theta\bigr] \bigl(Q_i^{a_i}(s_t) - Q^\ast\bigr).
\]
Since $(1-\alpha_i) + \alpha_i \theta \in [0, \lambda] \subset [0,1)$, the pre-rounding value lies on the same side of $Q^\ast$ as $Q_i^{a_i}(s_t)$ and at distance at most $\lambda d$ from $Q^\ast$. Rounding to the nearest grid point moves it by at most $\eta/2$, so $|Q_i^{a_i}(s_{t+1}) - Q^\ast| \leq \lambda d + \tfrac{\eta}{2}$.

\emph{Assumption~\ref{ax:contract} (outer region, $d > C\eta$).} Strict contraction requires $\lambda d + \tfrac{\eta}{2} < d$, i.e., $\tfrac{\eta}{2} < (1-\lambda) d = \alpha_i(1-\delta) d$. Since $d > C\eta$ and $C \geq \tfrac{1}{2\alpha_i(1-\delta)}$,
\[
    \alpha_i(1-\delta) d > \alpha_i(1-\delta) C\eta \geq \tfrac{\eta}{2},
\]
so $|Q_i^{a_i}(s_{t+1}) - Q^\ast| < d$. For the second requirement, either the pre-rounding value is more than $\eta/2$ from $Q^\ast$, so rounding by at most $\eta/2$ leaves it on the same side (no overshoot); or it is within $\eta/2$ of $Q^\ast$, so $|Q_i^{a_i}(s_{t+1}) - Q^\ast| \leq \eta \leq C\eta$, and the update lands in the band. Either way, Assumption~\ref{ax:contract} holds.

\emph{Assumption~\ref{ax:absorb} (band, $d \leq C\eta$).} Then
\[
    |Q_i^{a_i}(s_{t+1}) - Q^\ast| \leq \lambda C\eta + \tfrac{\eta}{2} 
    \leq C\eta
    \iff \tfrac{\eta}{2} \leq (1-\lambda) C\eta = \alpha_i(1-\delta) C\eta
    \iff C \geq \tfrac{1}{2\alpha_i(1-\delta)},
\]
which holds by assumption, so the update stays in the band.

\emph{Assumption~\ref{ax:monotone} (monotone entry).}
The pre-rounding update map $q \mapsto (1-\alpha_i)q+\alpha_i V_i$ is nondecreasing in $q$: its slope is $1-\alpha_i(1-\delta)$ when $a_i$ is the greedy action and $1-\alpha_i$ otherwise, both of which are positive. Rounding to the nearest grid point preserves monotonicity, so the discrete update map $q\mapsto q^+$ is also nondecreasing. A nondecreasing map on a finite ordered set cannot generate nontrivial cycles; hence, the iterates converge to a fixed point $q^\infty(q)$. Monotonicity implies that $q\geq \tilde q$ leads to $q^\infty(q)\geq q^\infty(\tilde q)$. Assumption~\ref{ax:monotone} holds.
\end{proof}

\subsection{Grid Approximation Assumptions}
\label{app:unaffected}

We impose the following auxiliary restrictions to make the discrete-grid approximation exact. First, the grid is assumed to be sufficiently fine.

\begin{ass}\label{ass:grid}
Denote $u(0,0)=\underline u$. The grid gap $\eta$ satisfies
\begin{equation}
    \eta<
    \min\left\{
    \min_{a\in A}\frac{u(a,a)-u(a-1,a-1)}{(2+\delta)C},\
    \min_{a>1} \max_{a'\in A}\frac{u(a',a)-u(a,a)}{2(1-\delta)C}
    \right\}.
\end{equation}
\end{ass}

We work on the following finite state space:
\begin{equation}
    \mathcal S=
    \left\{(\mathbf Q_1,\mathbf Q_2):
    Q_i^a\in[\underline Q^a,\bar Q]\cap\mathcal D,\ 
    a\in A,\ i\in I\right\},
\end{equation}
where $\underline Q^1=u(1,1)/(1-\delta)-C\eta$, $\underline Q^a=\underline u/(1-\delta)-C\eta$ for $a\neq1$, and $\bar Q\ge\max_{a,a'}u(a,a')/(1-\delta)$. This lower bound is lower than the one used in the main text by $C\eta$. The difference is natural, as updates are only required to stay within the $C\eta$-band around the fixed point.

Under the following two specifications, the main-text statements in
Proposition~\ref{prp:absorbing} and Theorem~\ref{thm:ss} hold exactly.

\begin{ass}\label{ass:aligned_grid}
Every value of the form $u(a_i,a_{-i})+\delta\,u(a_j,a_{-j})/(1-\delta)$, over all players and profiles $(a_i,a_{-i})$ and $(a_j,a_{-j})$, lies on the grid $\mathcal D$.
\end{ass}

\begin{ass}\label{ass:projection_path}
Fix a state $s_t$ and suppose the profile $(a_i, a_{-i})$ is played repeatedly with all other Q-values held fixed, so that the target $Q^\ast := Q^\ast_i(a_i, a_{-i}; s_t)$ stays constant. Then the iterates of $Q_i^{a_i}$ converge, after finitely many updates, to a grid point $\hat{q} \in \mathcal{D}$ with $|\hat{q} - Q^\ast| \leq C\eta$, and $\hat{q}$ is the unique fixed point of the update. If $Q^\ast \in \mathcal{D}$, then $\hat{q} = Q^\ast$.
\end{ass}

These two specifications are convenient approximations. When the grid is sufficiently fine, the band radius $C\eta$ is negligible, and requiring admissible updates to reach the fixed point exactly instead of allowing the value to remain anywhere within the band makes the argument cleaner. Under Assumption~\ref{ass:aligned_grid}, the fixed points that arise in the frozen dynamics lie on the grid. If $a_i$ is greedy, its fixed point is the long-run value $u(a_i, a_{-i})/(1-\delta)$. If $a_i$ is not greedy, its fixed point is $u(a_i, a_{-i}) + \delta \max_{a' \neq a_i} Q_i^{a'}(s_t)$; once the greedy action has reached its long-run value $u(a_j, a_{-j})/(1-\delta)$, this becomes $u(a_i, a_{-i}) + \delta\, u(a_j, a_{-j})/(1-\delta)$. Both forms lie in $\mathcal{D}$ by Assumption~\ref{ass:aligned_grid}, so Assumption~\ref{ass:projection_path} lets the discrete dynamics reach the relevant fixed point exactly.

\section{Proofs of the Theoretical Results}\label{app:proofs}

Throughout this section, updates are admissible and asynchronous, except in the last subsection, where they are synchronous, and the grid satisfies Assumption~\ref{ass:grid}. The unperturbed dynamics are induced by greedy play, as in \eqref{eq:unperturbed}, and the perturbed dynamics by exploration satisfying Assumption~\ref{ass:perturbed_dynamics}.

\subsection{Unperturbed Dynamics}

\begin{lemma}\label{lem:lower-bound}
For any recurrent class $S \in \mathcal{R}$, the following properties hold.
\begin{enumerate}
\item\label{item:convergence} For every greedy profile $(a_1,a_2)\in G(S)$, there exists a state $s\in S$ with $(a_1,a_2)\in G(s)$ and
\[
Q_i^{a_i}(s) \in \left[\frac{u(a_i,a_{-i})}{1-\delta} - C\eta,\
\frac{u(a_i,a_{-i})}{1-\delta} + C\eta \right]
\qquad\text{for all } i \in I.
\]
\item\label{item:lower-bound} There exist constants $(q_1^S,q_2^S)\in\mathbb R^2$ such that, for every $i\in I$ and every $s\in S$, any action used somewhere in $S$ but not greedy at $s$ attains the lower bound:
\[
Q_i^a(s)=q_i^S \qquad \text{for all } a\in G_i(S)\setminus G_i(s).
\]
Moreover, if player $i$ has more than one greedy action at $s$, then every action in $G_i(S)$ attains the lower bound at $s$:
\[
|G_i(s)|\ge 2 \quad\Longrightarrow\quad Q_i^a(s)=q_i^S \quad\text{for all }
a\in G_i(S).
\]
\end{enumerate}
\end{lemma}

\begin{proof}
\emph{Part \ref{item:convergence}.} Since $(a_1,a_2)\in G(S)$ is greedy, by the unperturbed dynamics \eqref{eq:unperturbed} and recurrency of $S$ it is played for arbitrarily many periods with positive probability, with each $a_i$ greedy so that $Q^\ast_i(a_i,a_{-i};s_t)\equiv (a_i,a_{-i})/(1-\delta)$ is constant. By Lemma~\ref{lemma:band-convergence}, the process reaches a state $s\in S$ with $(a_1,a_2)\in G(s)$ and $|Q_i^{a_i}(s)-u(a_i,a_{-i})/(1-\delta)|\le C\eta$ for
all $i$.

\emph{Part \ref{item:lower-bound}.} The argument follows Lemma~2 of
\cite{dolgopolov2024reinforcement}. Fix a player $i$ and suppose $|G_i(S)|\ge2$
(otherwise the claim is vacuous).

\emph{Step 1: the value of an action is the same when first and last played.}
Consider a state at which player $i$ switches to an action $a$, plays it for
several periods, and then switches away to another action $a''$; label the
switch-in state $s_1$ and the switch-out state $s_\tau$ (the first sequence in
Table~\ref{tab:sequences_action_transition}). We claim
$Q_i^{a}(s_1)=Q_i^{a}(s_\tau)$.

\begin{table}[H]
    \centering
    \begin{tabular}{c|ccccc|c|ccccc|c|cc}
         \hline
         state
         & $s_0$ & $s_1$ & $\cdots$ & $s_{\tau-1}$ & $s_\tau$ &$\cdots$ 
         & $s'_0$ & $s'_1$ & $\cdots$ & $s'_{\tau'-1}$ & $s'_{\tau'}$ &$\cdots$
         & $s_0$ & $s_1$ \\
         action of $i$
         & $a'$ & $a$ & $\cdots$ & $a$ & $a''$ &$\cdots$
         & $a'''$ & $a$ & $\cdots$ & $a$ & $a''''$ &$\cdots$
         & $a'$ & $a$ \\
         \hline
        \end{tabular}
    \caption{Sequences of action transitions. The first block is the episode
    from switch-in ($s_1$) to switch-out ($s_\tau$) of action $a$; the second
    block $s'_0,\dots,s'_{\tau'}$ is a later episode of $a$; the process then
    returns to $s_0,s_1$.}
    \label{tab:sequences_action_transition}
\end{table}

Between $s_1$ and $s_\tau$ only $a$ is played, so $Q_i^{a''}$ is unchanged: $Q_i^{a''}(s_1)=Q_i^{a''}(s_\tau)$. Since $a$ is greedy at $s_1$ and $a''$ is greedy at $s_\tau$,
\[
Q_i^{a}(s_1)\ge Q_i^{a''}(s_1)=Q_i^{a''}(s_\tau)\ge Q_i^{a}(s_\tau),
\]
so $Q_i^{a}(s_1)\ge Q_i^{a}(s_\tau)$. The same argument applies to \emph{every} episode in which $a$ is switched to, played, and then abandoned: the value of $a$ is weakly lower when abandoned than when first chosen. In other words, $Q_i^a$ can never strictly increase over such an episode.

Suppose, for contradiction, that $Q_i^{a}(s_1)>Q_i^{a}(s_\tau)$. As $S$ is recurrent, the process returns to $s_1$ with positive probability, restoring $Q_i^{a}=Q_i^{a}(s_1)>Q_i^{a}(s_\tau)$. Hence between $s_\tau$ and this return there is an episode in which $a$ is switched to and later abandoned while its value strictly increases, contradicting the previous paragraph. Therefore $Q_i^{a}(s_1)=Q_i^{a}(s_\tau)$, and combined with above inequalities, $Q_i^{a}(s_\tau)=Q_i^{a''}(s_\tau)$: at a switch, the abandoned and the newly chosen actions have equal Q-values.

\emph{Step 2: a common lower bound.}
Take a sequence of actions $k_1,k_2,\ldots,k_L$ that visits every action in $G_i(S)$ (repetitions allowed). Let $s_\ell$ denote a switching state at which player $i$ switches from $k_\ell$ to $k_{\ell+1}$, $\ell=1,\ldots,L-1$. By Step~1, we have
\[
    Q_i^{k_\ell}(s_\ell)=Q_i^{k_{\ell+1}}(s_\ell) = Q_i^{k_{\ell+1}}(s_{\ell+1}) = Q_i^{k_{\ell}}(s_{\ell+1}).
\]
Combining these equalities over $\ell$ shows that all switching values along the sequence coincide. Denote this common value by $q_i^S$. Since \eqref{upd:async}, an action is not updated while it is not played, so any action in $G_i(S)$ that is not greedy at a state $s$ retains this switching value. Therefore, $Q_i^{a}(s)=q_i^S$, for every $a\in G_i(S)\setminus G_i(s)$.

\emph{Step 3: the multi-greedy case.}
If $|G_i(s)|\ge2$, then some greedy action at $s$ is one that player $i$ is about to switch to or from; by Step~2 its value equals $q_i^S$, and hence, being tied with the other greedy actions at $s$, all actions in $G_i(S)$ have value $q_i^S$ at $s$.
\end{proof}

\begin{proposition}\label{prp:absorbing_prime}
For any game satisfying Assumption~\ref{ass:game}, a set of states
$S\subseteq\mathcal S$ is a recurrent class only if there exists $a\in A$ such that
$G(S)=\{(a,a)\}$ and, for every $i\in I$,
\begin{equation}
    Q_i^a(s) \in \left[\frac{u(a,a)}{1-\delta} -C\eta,\   \frac{u(a,a)}{1-\delta} +C\eta\right]
    \qquad\text{for all }s\in S.
\end{equation}
Conversely, for every $a\in A$, there exists a recurrent class
$S\subseteq\mathcal S$ with $G(S)=\{(a,a)\}$.
\end{proposition}

We first prove two claims.

\begin{claim}\label{cl:off-diagonal}
    No off-diagonal profile is played in any recurrent class: for any $S \in \mathcal{R}$, there is no $(a_i, a_{-i}) \in G(S)$ with $a_i \neq a_{-i}$.
\end{claim}

\begin{proof}
First, we show that $Q_i^1\ge u(1,1)/(1-\delta)-C\eta$ holds forever. By Assumption~\ref{ass:game}, $u(1,a)\ge u(1,1)$ for all $a$, so whenever action $1$ is updated its fixed point is at least $u(1,1)/(1-\delta)$. Hence under any admissible update, once $Q_i^1\ge u(1,1)/(1-\delta)-C\eta$ it remains so.

Now suppose, toward a contradiction, $(a_i,a_{-i})\in G(S)$ with $a_i>a_{-i}$ for some $S\in\mathcal{R}$.
By Lemma~\ref{lem:lower-bound}\ref{item:convergence}, there is a state $s\in S$
with $(a_i,a_{-i})\in G(s)$ and $Q_i^{a_i}(s)\le\underline{u}/(1-\delta)+C\eta$.
Since $Q_i^1(s)\ge u(1,1)/(1-\delta)-C\eta$, Assumption~\ref{ass:grid} gives
\[
    Q_i^1(s)>Q_i^{a_i}(s)\ge Q_i^{a'}(s)\qquad\text{for all }a'>1,
\]
so action $1$ is the unique greedy action for player $i$ at $s$. But then
$(a_i,a_{-i})$ with $a_i>1$ cannot be played, contradicting
$(a_i,a_{-i})\in G(s)$.
\end{proof}

\begin{claim}\label{cl:on-diagonal}
Each recurrent class contains exactly one symmetric profile: for any $S\in\mathcal{R}$, $G(S)=\{(a,a)\}$ for some $a\in A$.
\end{claim}

\begin{proof}
Suppose $(a,a),(a',a')\in G(S)$ with $a\neq a'$. By
Lemma~\ref{lem:lower-bound}\ref{item:convergence}, there are states
$s,s'\in S$ such that $(a,a)\in G(s)$, $(a',a')\in G(s')$, with $Q_i^a(s)$ in
the $C\eta$-band around $u(a,a)/(1-\delta)$ and $Q_i^{a'}(s')$ in the
$C\eta$-band around $u(a',a')/(1-\delta)$.

By Claim~\ref{cl:off-diagonal}, action $a$ is updated only when the opponent
also plays $a$, so its fixed point remains $u(a,a)/(1-\delta)$; once $Q_i^a$
enters its band, admissibility and asynchronicity keep it there throughout
$S$. Thus $Q_i^a(s')$ also lies in the band around $u(a,a)/(1-\delta)$. The same
argument gives that $Q_i^{a'}(s)$ lies in the band around
$u(a',a')/(1-\delta)$.

By Lemma~\ref{lem:lower-bound}\ref{item:lower-bound}, $q_i^S = Q_i^{a'}(s)$ and $q_i^S =Q_i^a(s')$. Hence the same value $q_i^S$ lies in both bands, contradicting Assumption~\ref{ass:grid}, which makes the two bands disjoint. Therefore $G(S)$ contains exactly one symmetric profile.
\end{proof}

Now we prove Proposition~\ref{prp:absorbing_prime}.

\begin{proof}

Let $S\subseteq\mathcal S$ be a recurrent class. By Claim~\ref{cl:on-diagonal},
there exists $a\in A$ such that $G(S)=\{(a,a)\}$. Since only $a$ is played in
$S$, every value $Q_i^{a'}$ with $a'\neq a$ is never updated and is therefore
constant across $S$. As both players keep selecting $a$,
Lemma~\ref{lem:lower-bound}\ref{item:convergence} implies $Q_i^a(s)$ in the $C\eta$-band around $u(a,a)/(1-\delta)$. This proves the necessary part.

Conversely, fix any $a\in A$. Consider the set of states in which $(a,a)$ is the greedy profile and, for each player $i$, $Q_i^a$ lies in the $C\eta$-band around $u(a,a)/(1-\delta)$, while all other Q-values are fixed and strictly lower than $u(a,a)/(1-\delta) - C\eta$ so that $a$ remains greedy. This set is nonempty by Assumption~\ref{ass:grid}. If $(a,a)$ is played, the values $Q_i^a$ remain in the band by \eqref{ax:absorb}, and the other Q-values are unchanged by \eqref{upd:async}. Hence the set is closed under the unperturbed
dynamics. Since the state space is finite, this closed set contains a recurrent
class $S$, and by construction $G(S)=\{(a,a)\}$. This proves the converse.
\end{proof}

\begin{corollary}\label{cor:exact}
Suppose the grid satisfies Assumption~\ref{ass:aligned_grid} and the updating rule satisfies Assumption~\ref{ass:projection_path}. Then Proposition~\ref{prp:absorbing} holds.
\end{corollary}

\begin{proof}
Let $S$ be a recurrent class. By Proposition~\ref{prp:absorbing_prime}, there exists $a\in A$ such that $G(S)=\{(a,a)\}$ and, for every $s\in S$ and $i\in I$, $Q_i^a(s)$ lies in the $C\eta$-band around $u(a,a)/(1-\delta)$. Moreover, all $Q_i^{a'}$ with $a'\neq a$ are never updated in $S$, and hence are constant throughout $S$. By Assumption~\ref{ass:aligned_grid}, $u(a,a)/(1-\delta)$ lies on the grid; by Assumption~\ref{ass:projection_path}, repeated admissible updates of $a$ reach this point in finitely many steps and keep it fixed. Therefore all states in $S$ have the same value $Q_i^a=u(a,a)/(1-\delta)$ and the same frozen values $Q_i^{a'}$, so $S$ is a singleton. Thus any recurrent class $\{s\}$ satisfies \eqref{eq:absorbing}, proving the ``only if'' direction.

Conversely, suppose $s$ satisfies \eqref{eq:absorbing}. Then $a$ is the unique greedy action for both players, so $(a,a)$ is played. Since $Q_i^a(s)=u(a,a)/(1-\delta)$ lies on the grid and is the fixed point of the updating under $(a,a)$, Assumption~\ref{ass:projection_path} keeps $Q_i^a$ unchanged under further updates of $a$. All other Q-values are unchanged by \eqref{upd:async}. Hence $s$ is absorbing.
\end{proof}

\subsection{Perturbed Dynamics and Stochastic Stability}

We first recall the tree characterization of stochastic stability due to \cite{kandori1993learning} and \cite{young1993evolution}. A family $\{P_\varepsilon\}$ is a \textbf{\textit{weakly regular perturbation}} of $P_0$ if, for small $\varepsilon>0$: (1) $P_\varepsilon$ is irreducible;\footnote{\cite{young1993evolution} additionally requires $P_\varepsilon$ to be aperiodic. Irreducibility alone, however, suffices for uniqueness of the stationary distribution, convergence, and the tree characterization in Theorem~\ref{thm:young}, which follows from the tree representation of stationary distributions in \citet[Lemma~3.1, p.~158]{freidlin1984random} and requires only irreducibility. Aperiodicity guarantees convergence of the distribution at each period to the stationary distribution; without it, the stationary distribution instead represents the long-run average fraction of time spent in each state \citep[see, e.g.,][]{kandori1995evolution}. Since stochastic stability concerns states that persist in the long run rather than period-by-period convergence, irreducibility is sufficient.}
(2) $P_\varepsilon(s,s')\to P_0(s,s')$ as $\varepsilon\to0$ for all$s,s'$; and (3) whenever $P_\varepsilon(s,s')>0$ for sufficiently small $\varepsilon$, there exists a cost $c(s,s')\geq0$ such that 
\begin{equation}
    0<\lim_{\varepsilon\to0}\varepsilon^{-c(s,s')}P_\varepsilon(s,s')<\infty.
\end{equation}

Following \citet{kandori1993learning} and \citet{young1993evolution}, we characterize stochastically stable states via the tree approach. Define the minimal cost of moving from state $s$ to state $s'$, possibly through intermediate states, as
\begin{equation}
    C(s,s') := \min_{(s_0 = s,\, s_1, \ldots, s_L = s')} 
    \sum_{\ell=0}^{L-1} c(s_\ell, s_{\ell+1}),
\end{equation}
where the minimum is over all paths from $s$ to $s'$. For recurrent classes $S, S' \in \mathcal{R}$, set
\begin{equation}
    C(S, S') := \min_{s \in S,\ s' \in S'} C(s, s').
\end{equation}
Since states within a recurrent class communicate at zero cost, the choice of representatives is immaterial: $C(S, S') = C(s, s')$ for any $s \in S$ and 
$s' \in S'$.

An \textbf{\textit{$S$-tree}} $h$ is a directed graph on $\mathcal{R}$ in which every class other than $S$ has exactly one outgoing edge and, following these edges from any class, one eventually reaches $S$. Its cost is the sum of its edge costs, $C(h) := \sum_{(S', S'') \in h} C(S', S'')$, where $(S', S'') \in h$ denotes a directed edge of $h$ from $S'$ to $S''$. Let
\begin{equation}
    C^\ast(S) := \min_{h \in H_S} C(h)
\end{equation}
denote the minimum cost among all $S$-trees, where $H_S$ is the set of $S$-trees.

\begin{theorem}[\citealp{young1993evolution}]\label{thm:young}
Let $P_0$ be a finite Markov chain with set of recurrent classes $\mathcal R$, and let $\{P_\varepsilon\}$ be a weakly regular perturbation of $P_0$. For each small $\varepsilon>0$, let $\mu_\varepsilon$ be the unique stationary distribution of $P_\varepsilon$. Then:
\begin{enumerate}
\item as $\varepsilon\to0$, $\mu_\varepsilon$ converges to a stationary distribution $\mu_0$ of $P_0$;
\item a state is stochastically stable, i.e.\ $\mu_0$ assigns it positive probability, if and only if it belongs to a recurrent class $S\in\mathcal R$ such that $C^\ast(S)\le C^\ast(S')$ for all $S'\in\mathcal R$.
\end{enumerate}
\end{theorem}

\paragraph{The Nash neighborhood.}
 Define $S^{NE}$ as the set of states $s$ satisfying
\begin{equation}
    \left|Q_i^1(s)-Q_i^1(s^{NE})\right|\le C\eta,
    \qquad
    \left|Q_i^a(s)-Q_i^a(s^{NE})\right|\le (1+\delta)C\eta
    \quad\text{for all }a > 1.
\end{equation}    
By Proposition~\ref{prp:absorbing_prime}, $S^{NE}$ contains at least one recurrent class of $P_0$. By Proposition~\ref{prp:reach-SNE} below shows that any $s \in \mathcal S$ can reach $S^{NE}$ under perturbed dynamics $P_\varepsilon$.

\subsubsection{The Nash-Proximity Order over Recurrent Classes}

For any recurrent class $S \in \mathcal{R}$, Proposition~\ref{prp:absorbing_prime} implies that $G(S) = \{(a(S), a(S))\}$ for a unique action $a(S)$. By \eqref{upd:async}, non-greedy actions are not updated within $S$, so for $a \neq a(S)$ we write $Q_i^a(S)$ for their constant value. For the greedy action, set $Q_i^{a(S)}(S) := \min_{s \in S} Q_i^{a(S)}(s)$.

The following definition and lemma extend Definition~\ref{def:order} and the first argument in Lemma~\ref{lem:property-order} of the main text from absorbing states to recurrent classes; the only change is that Q-values are evaluated at the class level, as defined above.

\begin{definition}\label{def:order_discrete}
Fix a mapping $\psi: A \to A$. Define a relation $\prec_\psi$ on $\mathcal{R}$ by $S \prec_\psi S'$ if one of the following holds:
\begin{enumerate}
\item\label{order_condition_1_d}
there exists $\hat{a} < \min\{a(S), a(S')\}$ with $Q_1^{\hat{a}}(S) < Q_1^{\hat{a}}(S')$ and $Q_1^a(S) = Q_1^a(S')$ for all $a < \hat{a}$;
\item\label{order_condition_2_d} condition~\ref{order_condition_1_d} fails and $a(S) < a(S')$;
\item\label{order_condition_3_d} condition~\ref{order_condition_1_d} fails, $a(S) = a(S') > 1$, and $Q_2^{\psi(a(S))}(S) > Q_2^{\psi(a(S'))}(S')$.
\end{enumerate}
\end{definition}

\begin{lemma}\label{lemma:transitivity_discrete}
The relation $\prec_\psi$ is irreflexive and transitive. The proof of Lemma~\ref{lemma:transitivity} applies verbatim, with states replaced by recurrent classes.
\end{lemma}

\begin{proof}
Irreflexivity is immediate. We prove transitivity. Let
$S\prec_\psi S'$ and $S'\prec_\psi S''$.

If $S\prec_\psi S'$ holds by condition~\ref{order_condition_1_d}, let $\hat a$ be the first action below $\min\{a(S),a(S')\}$ at which $Q_1^{\hat a}(S)<Q_1^{\hat a}(S')$, with equality for all lower actions. We claim that condition~\ref{order_condition_1_d} also gives $S\prec_\psi S''$. If $S'$ and $S''$ first differ at some lower action $\tilde a<\hat a$, then $Q_1^{\tilde a}(S)=Q_1^{\tilde a}(S')$ and the first difference between $S'$ and $S''$ makes $\tilde a<\min\{a(S),a(S'')\}$ a valid threshold for $S\prec_\psi S''$. If instead $S'$ and $S''$ are equal at all actions below $\hat a$, then $Q_1^{\hat a}(S')=Q_1^{\hat a}(S'')$, so $Q_1^{\hat a}(S)<Q_1^{\hat a}(S'')$. Moreover, $\hat a<\min\{a(S),a(S'')\}$: indeed, if $S'\prec_\psi S''$ is given by condition~\ref{order_condition_1_d}, its threshold is no smaller than $\hat a$ and is itself below $a(S'')$; if $S'\prec_\psi S''$ is given by condition~\ref{order_condition_2_d} or~\ref{order_condition_3_d}, then $a(S')\le a(S'')$ and $\hat a<a(S')$. Thus $\hat a$ is a valid threshold for $S\prec_\psi S''$. Hence $S\prec_\psi S''$ by condition~\ref{order_condition_1_d}.

If $S\prec_\psi S'$ holds by condition~\ref{order_condition_2_d}, then
$a(S)<a(S')$. If $S'\prec_\psi S''$ is given by condition~\ref{order_condition_2_d} or~\ref{order_condition_3_d}, then $a(S')\le a(S'')$, so $a(S)<a(S'')$ and condition~\ref{order_condition_2_d} gives $S\prec_\psi S''$. If instead $S'\prec_\psi S''$ is given by condition~\ref{order_condition_1_d}, then either that first difference occurs below $a(S)$, in which case condition~\ref{order_condition_1_d} applies to $S$ and $S''$, or it occurs above $a(S)$, in which case $a(S)<a(S'')$ and condition~\ref{order_condition_2_d} applies.

Finally, if $S\prec_\psi S'$ holds by condition~\ref{order_condition_3_d}, then $a(S)=a(S')$ and the lower-action Q-values relevant to conditions~\ref{order_condition_1_d} and~\ref{order_condition_2_d} coincide in the comparison. Thus whichever condition gives $S'\prec_\psi S''$ also gives $S\prec_\psi S''$; if condition~\ref{order_condition_3_d} applies again, transitivity follows from the strict ordering of the relevant $Q_2^{\psi(a(S))}$ values. Hence $\prec_\psi$ is transitive.
\end{proof}

\begin{proposition}\label{prp:reach-SNE}
For every recurrent class $S\in\mathcal R$ with $S\not\subseteq S^{NE}$, there exist recurrent classes $n_0,n_1,\ldots,n_p\in\mathcal R$ with $n_0=S$, $n_p\subseteq S^{NE}$, and $C(n_{\ell},n_{\ell+1})=1$  for every  $\ell=0,\ldots,p-1$. Conversely, if $S\subseteq S^{NE}$ and $S'\not\subseteq S^{NE}$ are recurrent classes, then $C(S,S')\ge2$.
\end{proposition}

To 
prove Proposition~\ref{prp:reach-SNE}, we first prove Lemma~\ref{lem:one_mutation_prime}, which is the discrete counterpart of the second argument in Lemma~\ref{lem:property-order} in the main text.

\begin{lemma}\label{lem:one_mutation_prime}
There exists $\psi:A\to A$ such that for every $S\in\mathcal{R}$ with $a(S)>1$, there is $S'\in\mathcal{R}$ with $S'\prec_\psi S$ and $C(S,S')=1$.
\end{lemma}

\begin{proof}
By Assumptions~\ref{ass:game}.\ref{assumption:3} and~\ref{ass:grid}, since $a(S)>1$ there exists $\tilde a<a(S)$ with
\[
    u(\tilde a,a(S))>u(a(S),a(S))+2(1-\delta)C\eta;
\]
set $\psi(a(S))=\tilde a$ (choosing arbitrarily if several such $\tilde a$ exist). Perturb player~$2$ once to choose $\tilde a$, and, with no further perturbation, let the process converge to a recurrent class $S'$; hence $C(S,S')=1$. If $a(S')=1$, then $S'\prec_\psi S$ by condition~\ref{order_condition_2_d}, so assume $a(S')\neq 1$.

We first establish an auxiliary result: if the lowest action played by player~$1$ along the path is not greedy at the next recurrent class $S'$, then its Q-value \emph{strictly} decreases with positive probability.

\begin{claim}\label{cl:lowest_action_decrease}
Starting at $S\in\mathcal{R}$, perturb player~$2$ once and allow no further perturbation. Then, with positive probability, the next recurrent class $S'$ satisfies either $a(S')=\underline a_1$ or $Q_1^{\underline a_1}(S')<Q_1^{\underline a_1}(S)$, where $\underline a_1$ is the lowest action played by player~$1$ on the path from $S$ to $S'$.
\end{claim}

\begin{proof}
Consider the \emph{lowest-action tie-breaking} (LATB) rule, under which a player, whenever she has more than one greedy action, chooses the lowest one with probability one. Under this rule Claim~\ref{cl:off-diagonal} still holds, since $1$ is the lowest action and only one action profile is played in the recurrent class (Claim~\ref{cl:on-diagonal}). Hence any recurrent set $S''$ reached under LATB has a greedy action $a(S'')$ with
\begin{equation*}
    \bigl|Q_i^{a(S'')}(S'')-u(a(S''),a(S''))/(1-\delta)\bigr|\le C\eta,
\end{equation*}
\begin{equation*}
    Q_i^{a'}(S'')<Q_i^{a(S'')}(S'')\ \text{for }a'<a(S''),
    \qquad
    Q_i^{a'}(S'')\le Q_i^{a(S'')}(S'')\ \text{for }a'>a(S'').
\end{equation*}
The only difference from a recurrent class of $P_0$ is that $Q_i^{a'}=Q_i^{a(S'')}$ is allowed for $a'>a(S'')$.

Suppose that, after the perturbation, the perturbation-free process under LATB converges to a set $\tilde S'$ with positive probability, and let $\underline a_1$ be the lowest action played by player~$1$ on the path. Since player~$1$ is not perturbed from $S$ to $\tilde S'$ and only played actions are updated, choosing $\underline a_1$ greedily and later abandoning it weakly decreases its Q-value; as $\underline a_1$ is the lowest action ever played, under LATB the decrease is strict. Hence either $a(\tilde S')=\underline a_1$ or $Q_1^{\underline a_1}(\tilde S')<Q_1^{\underline a_1}(S)$.

If $\tilde S'\in\mathcal{R}$, we are done. Otherwise some player~$i$ has, at $\tilde S'$, an action $a'>a(\tilde S')$ with $Q_i^{a'}(\tilde S')=Q_i^{a(\tilde S')}(\tilde S')$; let $i$ play $a'$ when it is greedy while the opponent keeps $a(\tilde S')$. The realized payoff is $\underline u$, so the fixed point is $\underline u+\delta Q_i^{a(\tilde S')}(\tilde S')$, and by Assumption~\ref{ass:grid},
\[
    \bigl|\underline u+\delta Q_i^{a(\tilde S')}(\tilde S')
    -Q_i^{a(\tilde S')}(\tilde S')\bigr|
    \ge u(1,1)-(1-\delta)C\eta-\underline u>C\eta.
\]
Thus $Q_i^{a'}$ strictly decreases after updating. This turns $\tilde S'$ into a recurrent class $S'\in\mathcal{R}$ with $a(S')=a(\tilde S')$ and $Q_1^{\underline a_1}(S')=Q_1^{\underline a_1}(\tilde S')$.
\end{proof}

We now compare $S'$ with $S$, distinguishing cases by $a(S')$.

\emph{Case 1: $a(S')>a(S)$.} Player~$2$ is perturbed to $\tilde a=\psi(a(S))$. If player~$1$ never played an action below $a(S)$, then player~$2$ faces only actions $a' \geq a(S)$, so by Assumption~\ref{ass:game}.\ref{assumption:4} the fixed point of $a(S)$ is at least $u(a(S),a(S))/(1-\delta)$. By Assumption~\ref{ax:monotone}, the Q-value of $a(S)$ therefore stays weakly above $Q_2^{a(S)}(S)$, even if it briefly leaves and re-enters the band. Since every action other than $\tilde a$ is left unchanged and remains below $Q_2^{a(S)}(S)$, player~$2$'s greedy action stays in $\{\tilde a, a(S)\}$, so $a(S') \in \{\tilde a, a(S)\}$, contradicting $a(S') > a(S)$. Player~$2$'s greedy action therefore remains in $\{\tilde a,a(S)\}$, so $a(S')\in\{\tilde a,a(S)\}$, contradicting $a(S')>a(S)$. Hence player~$1$ must have played some $\underline a_1<a(S)$, and Claim~\ref{cl:lowest_action_decrease} gives $Q_1^a(S')=Q_1^a(S)$ for all $a<\underline a_1$ and $Q_1^{\underline a_1}(S')<Q_1^{\underline a_1}(S)$. Thus $S'\prec_\psi S$ by condition~\ref{order_condition_1_d}.

\emph{Case 2: $a(S')<a(S)$.} If player~$1$ played some action below $a(S')$ on the path, then as in Case~1, Claim~\ref{cl:lowest_action_decrease} yields $S'\prec_\psi S$ by condition~\ref{order_condition_1_d}. Otherwise $Q_1^a(S)=Q_1^a(S')$ for all $a<a(S')$, so $a(S')<a(S)$ directly gives $S'\prec_\psi S$ by condition~\ref{order_condition_2_d}.

\emph{Case 3: $a(S')=a(S)$.} If player~$1$ played some action below $a(S)$, again $S'\prec_\psi S$ by condition~\ref{order_condition_1_d}. Otherwise, player~$2$ plays $\tilde a$ only against player~$1$'s actions $a'\geq a(S)$. $\tilde{a}$ Since $u(\tilde a,a')\ge u(\tilde a,a(S))>u(a(S),a(S))$ by Assumption~\ref{ass:game}.\ref{assumption:4}, the fixed point for $\tilde a$ exceeds
\[
    u(\tilde a,a(S))/(1-\delta)>u(a(S),a(S))/(1-\delta)+2C\eta
\]
by Assumption~\ref{ass:grid}. As $Q_2^{\tilde a}(S)<u(a(S),a(S))/(1-\delta)+C\eta$, the update raises it: $Q_2^{\tilde a}(S')>Q_2^{\tilde a}(S)$, so $S'\prec_\psi S$ by condition~\ref{order_condition_3_d}.
\end{proof}

We now prove Proposition~\ref{prp:reach-SNE}.

\begin{proof}
We first prove the existence of a cost-one path to $S^{NE}$, considering two cases.

\emph{Case 1: $a(S)=1$.}
By Proposition~\ref{prp:absorbing_prime}, $Q_i^1(S)$ lies in the $C\eta$-band around $u(1,1)/(1-\delta)$. Since $S\not\subseteq S^{NE}$, some non-greedy Q-values are not yet in the regions defining $S^{NE}$. We update them one at a time by single mutations. If player $i$ mutates to some $a'\neq1$ while the opponent plays $1$, the fixed point of $Q_i^{a'}$ is $\underline u+\delta Q_i^1(S)$, which lies in
\[
    \left[
    \underline u+\frac{\delta u(1,1)}{1-\delta}-\delta C\eta,\,
    \underline u+\frac{\delta u(1,1)}{1-\delta}+\delta C\eta
    \right].
\]
By Assumption~\ref{ass:grid}, this interval, enlarged by the admissible $C\eta$-band, lies strictly below $u(1,1)/(1-\delta)-C\eta$. Hence after such an update, action $1$ remains the unique greedy action for player $i$, and the process returns to a recurrent class with $a(S)=1$.

Repeating this one-mutation update for every non-greedy action of each player, the corresponding Q-values move toward their fixed points and, by finiteness of the grid, enter the required regions in finitely many updates. Thus after finitely many one-mutation steps the process reaches $S^{NE}$.

\emph{Case 2: $a(S)>1$.}
By Lemma~\ref{lem:one_mutation_prime}, there is a recurrent class $S'\prec_\psi S$ with $C(S,S')=1$. Repeating this argument gives a strictly descending sequence under $\prec_\psi$. Since $\prec_\psi$ is transitive and irreflexive on the finite set $\mathcal R$, the sequence must terminate. It can terminate only at a recurrent class with $a(S)=1$, which reaches $S^{NE}$ by Case 1. Hence the desired cost-one path to $S^{NE}$ exists.

For the converse, take any path from $S\subseteq S^{NE}$ to $S'\not\subseteq S^{NE}$. Along this path there is a first transition leaving $S^{NE}$. By the construction of $S^{NE}$ and Assumption~\ref{ass:grid}, such a transition cannot occur under a single mutation and therefore has cost at least $2$. Hence every path from $S$ to $S'$ has cost at least $2$, so $C(S,S')\ge2$.
\end{proof}

\subsubsection{Proof of Theorem~\ref{thm:ss}}

Let $\{\mathcal{S}_\varepsilon^\ell\}_\ell$ denote the recurrent classes of $P_\varepsilon$. By Assumption~\ref{ass:perturbed_dynamics}, $P_\varepsilon(s,s') > 0$ for some $\varepsilon \in (0,1)$ if and only if it holds for every $\varepsilon \in (0,1)$; hence each recurrent class $\mathcal{S}_\varepsilon^\ell$ does not depend on $\varepsilon$, and we write it $\mathcal{S}_+^\ell$. Each recurrent class carries its own stationary distribution; let $\mu_\varepsilon^\ell$ denote the stationary distribution of $P_\varepsilon$ supported on $\mathcal{S}_+^\ell$.

Unlike the classical setting, which requires the perturbed chain to have a unique stationary distribution, we allow $P_\varepsilon$ to admit multiple stationary distributions. The purpose of stochastic stability is to select a small subset of recurrent classes when the unperturbed process has many, although the selected states need not be unique. Under Assumptions~\ref{ass:aligned_grid} and~\ref{ass:projection_path}, however, every absorbing state can reach $s^{NE}$. Hence, with these additional assumptions imposed in the main text, $P_\varepsilon$ has a unique recurrent class and therefore a unique stationary distribution.

Since under both $P_0$ and $P_\varepsilon$ the process cannot leave $\mathcal{S}_+^\ell$ once it enters, we may restrict attention to $\mathcal{S}_+^\ell$. The restricted chain $\tilde{P}_\varepsilon^\ell$ is irreducible, since $\mathcal{S}_+^\ell$ is by definition a single recurrent class of $P_\varepsilon$. By Assumption~\ref{ass:perturbed_dynamics}, $\{\tilde{P}_\varepsilon^\ell\}$ is a weakly regular perturbation of $\tilde{P}_0^\ell$, so Theorem~\ref{thm:young} applies on each $\mathcal{S}_+^\ell$: as $\varepsilon \to 0$, $\mu_\varepsilon^\ell$ converges to a stationary distribution $\mu_0^\ell$ of $P_0$. Extending the concept in \cite{Foster1990}, we say that a state is \textbf{\textit{stochastically stable}} if it lies in the support of some $\mu_0^\ell$.

\begin{theorem}\label{thm:ss_prime}
A state $s \in \mathcal{S}$ is stochastically stable only if $s \in S^{NE}$.
\end{theorem}

\begin{proof}
Fix a recurrent class $\mathcal{S}_+^\ell$ of $P_\varepsilon$. By Proposition~\ref{prp:reach-SNE}, it contains a $P_0$-recurrent class in $S^{NE}$. If every $P_0$-recurrent class in $\mathcal{S}_+^\ell$ lies inside $S^{NE}$, there is nothing to prove. Thus suppose that at least one $P_0$-recurrent class in $\mathcal{S}_+^\ell$ lies outside $S^{NE}$. We show that for any outside $P_0$-recurrent class $S_k$, there exists a $P_0$-recurrent class $S_l^{NE}$ inside $S^{NE}$ with strictly lower cost $C^*$. Then, by Theorem~\ref{thm:young}, all stochastically stable states lie in $S^{NE}$.

Fix an outside $P_0$-recurrent class $S_k$ and a minimum-cost $S_k$-tree $h_k$, so $C^\ast(S_k)=C(h_k)$. Applying Lemma~\ref{lem:one_mutation_prime} repeatedly from $S_k$ produces a sequence
\[
    S_k=n_0\ \to\ n_1\ \to\ \cdots\ \to\ n_p=S^{NE}_l,
\]
where each edge $n_j\to n_{j+1}$ has cost $1$ and the endpoint $n_p=S^{NE}_l$ lies inside $S^{NE}$. Since $S_k\in\mathcal S_+$, every $n_j\in\mathcal S_+$. Form $h_l$ from $h_k$ by adding the path edges $n_0\to n_1\to\cdots\to n_p$ and deleting the original outgoing edge of each $n_0,\dots,n_p$ in $h_k$.

Then $h_l$ is a valid $S^{NE}_l$-tree. Every node except $S^{NE}_l$ has exactly one outgoing edge. Every node reaches $S^{NE}_l$: since $h_k$ is rooted at $n_0=S_k$, every node's $h_k$-path leads to $n_0$ and hence first meets some $n_j$; from there it follows the new path $n_j\to\cdots\to n_p=S^{NE}_l$.

Since the deleted edge out of $S^{NE}_l$ has cost at least $2$, whereas the new edge out of $S_k$ has cost $1$. Other deleted edge is replaced by an edge with the least cost $1$. Therefore, 
\[
    C^\ast(S^{NE}_l)\le C(h_l)\le C(h_k)-2+1=C^\ast(S_k)-1<C^\ast(S_k),
\]
then all states in $S_k$ cannot be stochastically stable.
\end{proof}

When the grid satisfies Assumption~\ref{ass:aligned_grid} and the updating rule satisfies Assumption~\ref{ass:projection_path}, Theorem~\ref{thm:ss} follows as a corollary of Theorem~\ref{thm:ss_prime}. Indeed, any recurrent class with $a(S)=1$ has $Q_i^1=u(1,1)/(1-\delta)$ exactly for each player $i$. The remaining non-greedy Q-values can then be brought to their fixed-point values one at a time through single mutations, so every state in $S^{NE}$ can reach $s^{NE}$ through a one-cost path as in Proposition~\ref{prp:reach-SNE}.

\subsection{Synchronous Updating}

\begin{ass}\label{ass:order_preserving}
    For any two actions $a, a'$ and any time $t$, if $Q_i^a(s_0)\geq Q_i^{a'}(s_0)$ and the fixed point of $a$ is weakly greater than that of $a'$ in every period up to $t$, then $Q_i^a(s_t)\geq Q_i^{a'}(s_t)$.\footnote{Since the fixed point in \eqref{eq:fixed_point} and the target in \eqref{eq:target} depend only on the payoff $u(a_i,a_{-i})$ given the same maximum Q-value, a weakly higher fixed point implies a weakly higher target.}
\end{ass}

\begin{ass}\label{ass:strict_order}
    Fix two actions $a, a'$ and an initial state $s_0$, and suppose the fixed point of $a$ is weakly greater than that of $a'$ in every period. If there exists $d > 0$ such that, for every $t$, the difference between the fixed points of $a$ and $a'$ is at least $d$ at some period $t' > t$, then $Q_i^a(s_\tau) \geq Q_i^{a'}(s_\tau)$ for some $\tau > 0$.
\end{ass}

It is easy to verify that standard Q-learning~\eqref{eq:synchronous} satisfies Assumptions~\ref{ass:order_preserving} and~\ref{ass:strict_order}, since it applies the same learning rate $\alpha_i$ to every action's Q-value. We now prove a general version of Proposition~\ref{prop:syn}.

\begin{proposition}\label{prp:synchronous}
    Under admissible synchronous updating (Definition~\ref{def:admissible}) and Assumptions~\ref{ass:order_preserving} and~\ref{ass:strict_order}, every recurrent class $S$ of the unperturbed dynamics satisfies $G(S) = \{(1,1)\}$.
\end{proposition}

\begin{proof}
Fix any recurrent class $S$, define 
\[
    \overline{a}(S) := \min\bigl\{ \max(a_1, a_2) : (a_1, a_2) \in G(S) \bigr\},
\]
Let $s\in S$ such that $(a_1,a_2)\in G(s)$ with $a_i = \overline{a}(S)$ and $a_i\ge a_{-i}$ for some $i\in\{1,2\}$. Hence $\min G_{-i}(s)\le \overline{a}(S)$. If $\min G_{-i}(s)< \overline{a}(S)$ , then $\min G_{i}(s) = \overline{a}(S) $. Therefore, $\exists j\in \{1,2\}$ such that $\min G_j (s) = \overline{a}(S)$. Without loss of generality, set $j=1$.

Start from $s_0 = s$, we adopt lowest-action tie-breaking (LATB) for some finite periods. 

First, consider the case $\overline{a}(S) = 1$, so that $\min G_2(s_0) = 1$. By Assumption~\ref{ass:game} and \ref{ass:order_preserving}, both players play action $1$ throughout the LATB phase. By Lemma~\ref{lemma:band-convergence}, once both play action $1$, the Q-value of every other action converges to at most
\[
    \underline{u} + \frac{\delta\, u(1,1)}{1-\delta} + C\eta 
    < \frac{u(1,1)}{1-\delta} - C\eta \leq Q_i^1,
\]
so action $1$ becomes strictly greedy. Hence, under the LATB rule for the first $\tau$ periods and any tie-breaking rule thereafter, the dynamics converge to a single state, so $S$ is a singleton with $G(S) = \{(1,1)\}$.

It remains to prove that it cannot be that $\overline{a}(S) > 1$, which is proved by contradiction. 

\begin{claim}\label{cl:exclude-up}
    During the $\tau$-period LATB phase, $\min G_i(s_t) \leq \overline{a}(S)$ for every $i \in \{1,2\}$ and $t \leq \tau$.
\end{claim}

\begin{proof}
    Suppose not, and let $\underline{t}$ be the first period at which $\min G_j(s_{\underline{t}}) > \overline{a}(S)$ for some player $j$. Then for every $t < \underline{t}$ and every player, the played action is at most $\overline{a}(S)$. Compare, for player $j$, the actions 
    $a := \min G_j(s_0) \leq \overline{a}(S)$ and 
    $a' := \min G_j(s_{\underline{t}}) > \overline{a}(S)$, so $a < a'$. By 
    Assumption~\ref{ass:game}.\ref{assumption:4}, the fixed point of $a$ is weakly above that of $a'$ in every period $t < \underline{t}$:
    \[
        Q^*_j(a, \min G_{-j}(s_t); s_t) 
        \geq Q^*_j(a', \min G_{-j}(s_t); s_t).
    \]
    Since also $Q_j^{a}(s_0) \geq Q_j^{a'}(s_0)$, Assumption~\ref{ass:order_preserving} gives $Q_j^{a}(s_{\underline{t}}) \geq Q_j^{a'}(s_{\underline{t}})$. But $a' = \min G_j(s_{\underline{t}})$ is greedy at $s_{\underline{t}}$ while $a$ is not, so $Q_j^{a'}(s_{\underline{t}}) > Q_j^{a}(s_{\underline{t}})$, a contradiction.
\end{proof}

\begin{claim}\label{cl:alternating}
    For any $\tau$, there exists $t > \tau$ such that, if the LATB is adopted for at least $t$ periods, then $\min G_2(s_t) = \overline{a}(S)$.
\end{claim}

\begin{proof}
    Suppose not: there exists $\hat \tau$ such that $\min G_2(s_t) < \overline{a}(S)$ for all $t > \hat \tau$. Since $S$ is a recurrent class, by the definition of $\overline{a}(S)$ and Claim~\ref{cl:exclude-up}, $\min G_1(s_t) = \overline{a}(S)$ for all $t > \hat \tau$. Comparing player~$1$'s actions $1$ and $\overline{a}(S)$, their fixed points satisfy, for every $t > \hat \tau$,
    \[
        Q_1^*\bigl(1, \min G_2(s_t); s_t\bigr) 
        - Q_1^*\bigl(\overline{a}(S), \min G_2(s_t); s_t\bigr)
        \geq \min_{a} u(1, a) - \underline{u} > 0,
    \]
    since action $1$ wins against any $\min G_2(s_t) < \overline{a}(S)$ while $\overline{a}(S)$ loses and earns $\underline{u}$. Thus the fixed point of action $1$ exceeds that of $\overline{a}(S)$ by a fixed positive amount infinitely often, so Assumption~\ref{ass:strict_order} gives some $\hat{t} > \hat{\tau}$ with $Q_1^{1}(s_{\hat t}) \geq Q_1^{\overline{a}(S)}(s_{\hat t})$. Under the LATB, player $1$ then plays action $1 < \overline{a}(S)$ at $s_{\hat t}$, contradicting $\min G_1(s_{\hat t}) = \overline{a}(S)$.
\end{proof}

\begin{claim}\label{cl:1-to-lower}
    There exists $\tau$ such that, if the LATB is adopted for $\tau' > \tau$ periods, then $\min G_1(s_t) < \overline{a}(S)$ for all $t \in [\tau, \tau']$.
\end{claim}

\begin{proof}
    Let $BR(\overline{a}(S))$ be one of the best responses to $\overline{a}(S)$; by Assumption~\ref{ass:game}, $BR(\overline{a}(S)) < \overline{a}(S)$. By Claim~\ref{cl:exclude-up}, the opponent plays at most $\overline{a}(S)$ throughout this phase, so the fixed point of $BR(\overline{a}(S))$ is weakly above that of $\overline{a}(S)$ in every period. By Claim~\ref{cl:alternating}, $\min G_2(s_t) = \overline{a}(S)$ for arbitrarily large $t$, and at each such period the two fixed points differ by a fixed positive amount:
    \[
        Q_1^*\bigl(BR(\overline{a}(S)), \overline{a}(S); s_t\bigr) 
        - Q_1^*\bigl(\overline{a}(S), \overline{a}(S); s_t\bigr)
        \geq u\bigl(BR(\overline{a}(S)), \overline{a}(S)\bigr) 
        - u\bigl(\overline{a}(S), \overline{a}(S)\bigr) > 0.
    \]
    By Assumption~\ref{ass:strict_order}, there is a period $\tau$ with $Q_1^{BR(\overline{a}(S))}(s_{\tau}) \geq Q_1^{\overline{a}(S)}(s_{\tau})$, and by Assumption~\ref{ass:order_preserving} this persists:
    \[
        Q_1^{BR(\overline{a}(S))}(s_t) \geq Q_1^{\overline{a}(S)}(s_t)
        \qquad \text{for all } t \geq \tau.
    \]
    Under LATB, player $1$  never plays $\overline{a}(S)$ after $\tau$, so $\min G_1(s_t) < \overline{a}(S)$ for all $t \in [\tau, \tau']$.
\end{proof}

Now, we show that $ \overline{a}(S) > 1$ cannot hold. By Claim~\ref{cl:1-to-lower}, there is $\tau$ such that $\min G_1(s_t) < \overline{a}(S)$ for all $t \geq \tau$. Then, for every such $t$, player $1$ plays some action $a_{1,t} < \overline{a}(S)$, so
\[
    Q_2^*\bigl(BR(\overline{a}(S)), a_{1,t}; s_t\bigr) 
    - Q_2^*\bigl(\overline{a}(S), a_{1,t}; s_t\bigr)
    \geq u\bigl(BR(\overline{a}(S)), a_{1,t}\bigr) 
    - u\bigl(\overline{a}(S), a_{1,t}\bigr) > 0,
\]
that is, the fixed point of $BR(\overline{a}(S))$ exceeds that of $\overline{a}(S)$ by a fixed positive amount. By Assumption~\ref{ass:strict_order} there is a period $\tau'$ with $Q_2^{BR(\overline{a}(S))}(s_{\tau'}) \geq Q_2^{\overline{a}(S)}(s_{\tau'})$, and by Assumption~\ref{ass:order_preserving} this persists, so under LATB player $2$ never plays $\overline{a}(S)$ after $\tau'$. Hence $\min G_2(s_t) < \overline{a}(S)$ for all $t \geq \tau'$ as well.

Therefore, at any $s_t \in S$ with $t \geq \tau'$, both players play below $\overline{a}(S)$, so $\max_i \min G_i(s_t) < \overline{a}(S)$, contradicting the definition of $\overline{a}(S)$.
\end{proof}

\section{Supplementary of the Simulation Part}

\subsection{Proof of Proposition~\ref{prp:hitting_time}}
\label{appendix:random_walk}

\begin{proof}
Let $T_l$ denote the expected time to reach the absorbing state $L$ from state $l \in \{0, 1, \ldots, L\}$, with $T_L = 0$, so that $H(L,q,p) = T_0$. Define the incremental time $d_l := T_l - T_{l+1}$ for $l \in \{0, \ldots, L-1\}$, i.e., the expected time to advance from $l$ to $l+1$.

For interior states $l \in \{1, \ldots, L-1\}$, first-step analysis gives
\[
    T_l = 1 + p\, T_{l+1} + q\, T_{l-1} + (1 - p - q)\, T_l,
\]
which rearranges to
\begin{equation}\label{eq:app_d_rec}
    d_l = \frac{1}{p} + \rho\, d_{l-1}, 
    \qquad \rho := \frac{q}{p}.
\end{equation}
At the reflecting boundary, $T_0 = 1 + p\, T_1 + (1-p)\, T_0$, which yields the initial condition $d_0 = 1/p$.

\emph{Case $p \neq q$ ($\rho \neq 1$).} Iterating \eqref{eq:app_d_rec} from $d_0 = 1/p$,
\[
    d_l = \frac{1}{p} \sum_{m=0}^{l} \rho^m 
    = \frac{1}{p} \cdot \frac{1 - \rho^{l+1}}{1 - \rho}
    = \frac{1 - \rho^{l+1}}{p - q},
\]
where the last equality uses $p(1 - \rho) = p - q$. Summing the increments,
\[
    H(L,q,p) = T_0 = \sum_{l=0}^{L-1} d_l
    = \frac{1}{p - q} \left( L - \rho\, \frac{1 - \rho^{L}}{1 - \rho} \right)
    = \frac{L}{p - q} 
    + \frac{q}{(p-q)^2} \left[ \left(\frac{q}{p}\right)^{L} - 1 \right],
\]
where the last equality uses $\rho/(1-\rho) = q/(p-q)$.

\emph{Case $p = q$ ($\rho = 1$).} Recursion \eqref{eq:app_d_rec} gives $d_l = d_{l-1} + 1/p$, so $d_l = (l+1)/p$, and
\[
    H(L,q,p) = \sum_{l=0}^{L-1} \frac{l+1}{p} = \frac{L(L+1)}{2p}. 
\]
\end{proof}

\subsection{Simulation Design}\label{Appendix:experiment_design}

\paragraph{Economic Environment.}
We conduct simulations in a Bertrand duopoly model. Two firms $i \in I = \{0, 1\}$ with marginal costs normalized to zero sell homogeneous goods and set prices via Q-learning. The consumer's value is normalized to one. Firm $i$ sets price $p_i \in A$; consumers buy from the cheaper firm, and ties split demand equally:
\begin{equation}
    q_i(p_i, p_j) = \begin{cases} 
        1 & \text{if } p_i < p_j \text{ and } p_i \leq 1,\\
        \frac{1}{2} & \text{if } p_i = p_j \text{ and } p_i \leq 1,\\
        0 & \text{otherwise}.
    \end{cases}
\end{equation}
Firm $i$'s payoff is $u_i(p_i, p_j) = p_i\, q_i(p_i, p_j)$.

In the baseline, $K=100$ and prices are uniformly spaced between $0.1$ and $1$. The price $0.1$ is the unique NE and it is strict.

\paragraph{Initialization.}
Since our theory predicts that the dynamics converge to the NE in the long run, we adopt the optimistic initialization used in \cite{asker2023impact, banchioartificial}, which makes convergence to the NE more difficult. Specifically, we initialize all Q-values to the highest feasible discounted payoff:
\[
    \max_{a,a'}\frac{u(a,a')}{1-\delta}.
\]

\paragraph{Exploration.}

We run simulations under a small, constant exploration rate $\varepsilon$. Since the horizon comprises $T/\varepsilon$ periods, direct period-by-period simulation is infeasible for small $\varepsilon$, and we use the following accelerated procedure.

We treat each block of $1/\varepsilon$ consecutive periods as one \emph{unit of time}. At the start of each unit, we pre-generate the exploration times of both agents: for each agent, the number of explorations is drawn from a Poisson distribution with mean one---which approximates the Binomial $B(1/\varepsilon, \varepsilon)$ for small $\varepsilon$---and the explorations are placed uniformly at random within the unit. All exploration times are then sorted chronologically.

Between two consecutive explorations, the dynamics are deterministic, and we compute them in closed form rather than period by period. Immediately after an exploration, we update the Q-values step by step until the greedy action $a$ stabilizes and the fixed point $u(a,a)/(1-\delta)$ exceeds the second-highest Q-value; from then on, the agent repeats $a$ deterministically until the next exploration. Let $\tau$ denote the number of remaining periods until that exploration. Over these $\tau$ periods, the profile $(a, a)$ is played with $a$ greedy, so the continuation term of the update is $\max_{a'} Q_i^{a'} = Q_i^a$, and the update reduces to a geometric contraction toward $u(a,a)/(1-\delta)$. Iterating gives the closed form
\begin{equation}\label{eq:accelerated_update}
    Q_i^a(s_{t+\tau}) 
    = (1 - \alpha_1)^{\tau}\, Q_i^a(s_t) 
    + \bigl[1 - (1 - \alpha_1)^{\tau}\bigr]\, \frac{u(a,a)}{1-\delta},
\end{equation}
where $\alpha_1$ denotes the effective learning rate in the two-rate decomposition \eqref{eq:two_learning_rate} and equals $\alpha(1-\delta)$ in standard Q-learning. We skip these $\tau$ periods and apply \eqref{eq:accelerated_update} directly. This preserves the exact dynamics while making long horizons computationally feasible.

\subsection{Data and Estimation Details}
\label{app:estimation}

This subsection describes the data and the procedure used to estimate the latent random-walk parameters. Throughout, we follow the notation introduced there: $i$ indexes the effective learning rate $\alpha_1$, $j$ indexes the explorative learning rate $\alpha_2$, and $e$ indexes the exploration profile. A \emph{series} is a pair $(j,e)$, i.e.\ a fixed value of $\alpha_2$ together with an exploration profile; within a series, the rows are indexed by $i$ (equivalently, by $\alpha_1$).

\subsubsection{Data}
\label{app:data}

\begin{table}[!t]
\centering
\caption{Descriptive statistics of convergence times}
\label{tab:desc}
\begin{threeparttable}
\begin{tabular}{lrrrrrrr}
\toprule
\multicolumn{8}{l}{\emph{Panel A: by exploration profile}} \\
\midrule
Profile & \# miss. & Miss. & $T_{\min}$ & $T_{\text{med}}$ & $T_{\max}$ & $\mathrm{ave}(\log T)$ & $\mathrm{sd}(\log T)$ \\
\midrule
\texttt{rr} & 71 & 7\%  & $2.6\times10^{2}$ & $1.7\times10^{3}$ & $6.1\times10^{9}$ & 8.45 & 3.08 \\
\texttt{rb} & 0  & 0\%  & $1.6\times10^{1}$ & $5.7\times10^{1}$ & $5.1\times10^{3}$ & 4.41 & 1.25 \\
\texttt{re} & 0  & 0\%  & $1.1\times10^{2}$ & $7.0\times10^{2}$ & $9.3\times10^{4}$ & 6.93 & 1.57 \\
\texttt{rn} & 0  & 0\%  & $5.0\times10^{2}$ & $2.1\times10^{3}$ & $1.4\times10^{5}$ & 8.00 & 1.28 \\
\texttt{bb} & 0  & 0\%  & $4.9\times10^{1}$ & $5.6\times10^{2}$ & $8.4\times10^{4}$ & 6.56 & 1.64 \\
\texttt{be} & 0  & 0\%  & $1.6\times10^{1}$ & $5.6\times10^{1}$ & $4.7\times10^{3}$ & 4.38 & 1.23 \\
\texttt{bn} & 0  & 0\%  & $2.7\times10^{0}$ & $6.0\times10^{1}$ & $3.8\times10^{3}$ & 4.38 & 1.24 \\
\texttt{ee} & 0  & 0\%  & $8.3\times10^{1}$ & $6.0\times10^{2}$ & $1.2\times10^{5}$ & 6.77 & 1.61 \\
\texttt{en} & 0  & 0\%  & $1.9\times10^{2}$ & $8.7\times10^{2}$ & $1.2\times10^{5}$ & 7.14 & 1.41 \\
\midrule
\multicolumn{8}{l}{\emph{Panel B: \texttt{rr} by $\alpha_2$}} \\
\midrule
$\alpha_2$ & $\alpha_1^{\min}$ & Miss. & $T_{\min}$ & $T_{\text{med}}$ & $T_{\max}$ & $\mathrm{ave}(\log T)$ & $\mathrm{sd}(\log T)$ \\
\midrule
$0.10$ & 0.017 & 12\% & $2.6\times10^{2}$ & $1.8\times10^{3}$ & $6.0\times10^{9}$ & 8.90 & 3.75 \\
$0.09$ & 0.017 & 12\% & $3.0\times10^{2}$ & $1.7\times10^{3}$ & $3.0\times10^{9}$ & 8.79 & 3.52 \\
$0.08$ & 0.016 & 11\% & $3.0\times10^{2}$ & $1.8\times10^{3}$ & $4.0\times10^{9}$ & 8.78 & 3.54 \\
$0.07$ & 0.015 & 10\% & $2.8\times10^{2}$ & $1.8\times10^{3}$ & $5.5\times10^{9}$ & 8.74 & 3.54 \\
$0.06$ & 0.014 & 9\%  & $2.9\times10^{2}$ & $1.8\times10^{3}$ & $6.1\times10^{9}$ & 8.65 & 3.38 \\
$0.05$ & 0.013 & 8\%  & $3.1\times10^{2}$ & $1.8\times10^{3}$ & $1.9\times10^{9}$ & 8.46 & 3.08 \\
$0.04$ & 0.011 & 6\%  & $3.4\times10^{2}$ & $1.7\times10^{3}$ & $2.2\times10^{9}$ & 8.35 & 2.92 \\
$0.03$ & 0.008 & 3\%  & $3.3\times10^{2}$ & $1.6\times10^{3}$ & $6.1\times10^{9}$ & 8.26 & 2.81 \\
$0.02$ & 0.005 & 0\%  & $3.4\times10^{2}$ & $1.7\times10^{3}$ & $1.2\times10^{9}$ & 8.05 & 2.31 \\
$0.01$ & 0.005 & 0\%  & $3.2\times10^{2}$ & $1.6\times10^{3}$ & $1.0\times10^{5}$ & 7.70 & 1.41 \\
\bottomrule
\end{tabular}
\begin{tablenotes}
\small
\item \textit{Note:}
Panel A aggregates the ten series $(j,e)$ sharing exploration profile $e$ (one per value of $\alpha_2$). 
Panel B reports the ten \texttt{rr} series separately, with $\alpha_1^{\min}$ the smallest effective learning rate at which convergence is recorded. `\# miss.' and `Miss.' are the number and share of non-converged cells. $T$ statistics are on the original scale; $\log T$ statistics on the log scale.
\end{tablenotes}
\end{threeparttable}
\end{table}

For each triple $(i,j,e)$ we run the two-agent Q-learning simulation and record the average convergence time $T_{ije}$. We set $\delta=0$ in all simulations. The learning-rate grid is
\[
\alpha_1 \in \{0.005, 0.006, \dots, 0.100\}
\quad(\text{step } 0.001),
\qquad
\alpha_2 \in \{0.01, 0.02, \dots, 0.10\}
\quad(\text{step } 0.01),
\]
and $e$ ranges over the nine exploration profiles
$\mathcal{E}=\{\texttt{rr}, \texttt{rb}, \texttt{rp}, \texttt{rn}, \texttt{bb},
\texttt{bp}, \texttt{bn}, \texttt{pp}, \texttt{pn}\}$.

\paragraph{Missing Cells.}
Under the \texttt{rr} mode, when $\alpha_1$ is small and $\alpha_2$ is large, the convergence time grows super-exponentially in $L$ and exceeds our computational budget. We cap each run at the order of $10^{10}$ periods; runs that do not converge within this limit are treated as missing and excluded from the estimation. As a result, larger $\alpha_2$ values leave fewer valid $\alpha_1$ values in the \texttt{rr} series. Table~\ref{tab:desc} summarizes the missing cells. Panel A shows that convergence is recorded in every cell except under the \texttt{rr} mode, where $7\%$ of cells fail to converge within the budget. Panel B further breaks down the \texttt{rr} series by $\alpha_2$: as $\alpha_2$ decreases, the missing share falls from $12\%$ to zero, while the smallest converging effective learning rate $\alpha_1^{\min}$ declines from $0.017$ to $0.005$. Thus, lower explorative learning rate allows convergence over a wider range of $\alpha_1$ values.

\paragraph{Convergence Times.}
As shown in Table~\ref{tab:desc}, the \texttt{rr} mode stands out sharply. It has by far the largest mean and dispersion of $\log T$, and its maximum convergence time reaches $6.1\times10^{9}$, several orders of magnitude above those of the other profiles. Within the \texttt{rr} mode, $\mathrm{ave}(\log T)$ decreases monotonically from $8.90$ to $7.70$ as $\alpha_2$ declines.

\subsubsection{Estimation}
\label{app:model}

By Assumptions~\ref{ass:length} and~\ref{ass:prob}, the length depends only on $\alpha_1$, $L_i = L(\alpha_1)$, and the transition probabilities depend only on $\alpha_2$ and the exploration profile, $(p_{je}, q_{je})$. The predicted convergence time is given by the hitting-time formula of Proposition~\ref{prp:hitting_time},
\[
\widehat{T}_{ije} = H\bigl(L_i, p_{je}, q_{je}\bigr).
\]
We first estimate the model directly and then impose the structural specification. Table~\ref{tab:fit} shows that both approaches provide a good fit.

\begin{table}[h]
\centering
\caption{Fit of the random-walk model}
\label{tab:fit}
\begin{threeparttable}
\begin{tabular}{lc}
\toprule
 & $R^2$ (log scale) \\
\midrule
Direct Estimation & $0.994$ \\
Structural Specification & $0.993$ \\
\bottomrule
\end{tabular}
\end{threeparttable}
\end{table}

\paragraph{Part 1: Direct Estimation.}
In the first part we estimate the latent parameters directly, imposing no functional form beyond Assumptions~\ref{ass:length} and~\ref{ass:prob}. We treat each length $L_i$ and each pair of transition probabilities $(p_{je}, q_{je})$ as free parameters. Because convergence times span many orders of magnitude, we fit on the log scale. For each valid cell we define the normalized log-residual
\[
\varepsilon_{ije}
= \frac{\log \widehat{T}_{ije} - \log T_{ije}}{s_{je}},
\]
where $s_{je}$ is the standard deviation of $\log T$ within series $(j,e)$, computed over its valid cells. This normalization places all series on a common scale, so that series with a larger dispersion of $\log T$ do not dominate the objective. We then solve
\begin{equation}\label{eq:res}
    \min_{\{L_i\},\,\{p_{je},\, q_{je}\}}
    \;\sum_{(j,e)} \sum_{i}
    \varepsilon_{ije}^{\,2},
\end{equation}
subject to $L_i$ decreasing in $\alpha_1$ (Assumption~\ref{ass:length}), $p_{je}, q_{je}\ge 0$, and $p_{je}+q_{je}\le 1$. This yields a nonparametric estimate of the length profile $\{L_i\}$ and of the transition probabilities for every series.

Table~\ref{tab:rw_structure} reports the estimated structure of the random-walk model based on directly estimated parameters. The length $L$ is almost perfectly log-log linear in the effective learning rate $\alpha_1$, with a slope of $-1.56$ and an $R^2$ of $0.999$. Under the \texttt{rr} mode, the forward and backward transition probabilities are well approximated by linear functions of the exploration learning rate $\alpha_2$, with the backward probability rising more steeply.

\begin{table}[h]
\centering
\caption{Estimated structure of the random-walk model}
\label{tab:rw_structure}
\begin{threeparttable}
\begin{tabular}{lcccc}
\toprule
 & Intercept & Slope & & $R^2$ \\
\midrule
$\log L = a + b\,\log\alpha_1$ & $-2.154$ & $-1.563$ & & $0.999$ \\
$p^{\texttt{rr}} = p_0 + p_1\,\alpha_2$ & $0.037$ & $0.347$ & & $0.750$ \\
$q^{\texttt{rr}} = q_0 + q_1\,\alpha_2$ & $0.031$ & $0.609$ & & $0.882$ \\
\bottomrule
\end{tabular}
\begin{tablenotes}
\small
\item \textit{Note:} The first row reports the log-log regression of the length $L$ on the effective learning rate $\alpha_1$. The second and third rows report the linear regressions of the forward and backward transition probabilities on the exploration learning rate $\alpha_2$ under \texttt{rr}.
\end{tablenotes}
\end{threeparttable}
\end{table}

\paragraph{Part 2: Structural Specification.}
Building on the structure identified above, we impose a parametric specification to extrapolate the convergence time beyond the observed grid. The direct estimates are defined only on the observed grid and therefore cannot predict convergence times at unobserved parameter values. This extrapolation is particularly useful for the \texttt{rr} mode, where we trace convergence time as a function of $(\alpha,\delta)$. We specify parametric forms for the two components that vary along the grid: the length and the transition probabilities of the \texttt{rr} random walk.

For the length, we assume a power-law relationship with the effective learning rate:
\[
L(\alpha_1)=\exp\bigl(a+b\log\alpha_1\bigr),
\qquad b<0.
\]
For the \texttt{rr} mode, we specify the forward and backward transition probabilities as linear functions of the exploratory learning rate:
\[
p^{\texttt{rr}}(\alpha_2)=p_0+p_1\alpha_2,
\qquad
q^{\texttt{rr}}(\alpha_2)=q_0+q_1\alpha_2.
\] 
Combining these two structural components with the hitting-time formula yields a smooth prediction of convergence time over the entire $(\alpha,\delta)$ space. The structural parameters are estimated by minimizing the log-residual in \eqref{eq:res}.

\begin{figure}[h]
    \centering
    \subfigure[Observed vs. Fitted]{
        \includegraphics[width=0.49\textwidth]{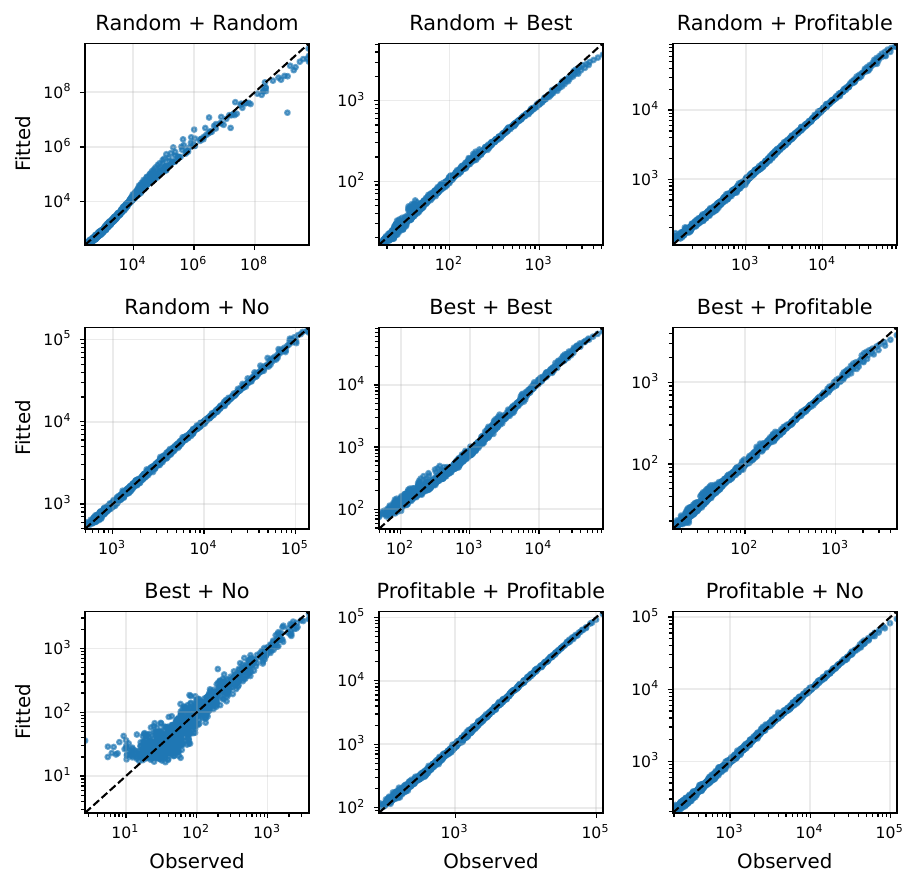}%
        \label{fig:linear_obs_fit}}
    \hfill
    \subfigure[Drift]{
        \includegraphics[width=0.49\textwidth]{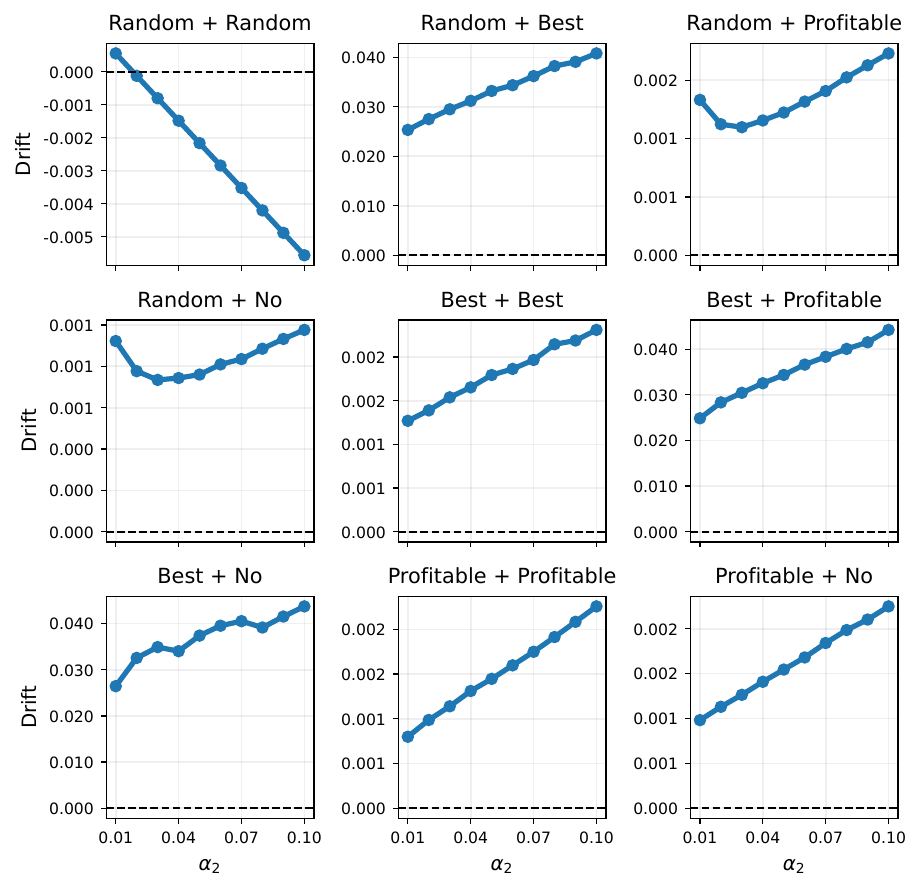}%
        \label{fig:linear_drift}}

    \caption{Fit of the random-walk model under structural specification.}
    \label{fig:linear_fit}
\end{figure}

Figure~\ref{fig:linear_fit} shows the fit of the structural specification and the estimated drift. Despite some differences in magnitude, the estimated drift exhibits the same pattern as under direct estimation.

\end{document}